\documentclass[sn-mathphys,Numbered]{sn-jnl}
\usepackage{graphicx}%
\usepackage{multirow}%
\usepackage{amsmath,amssymb,amsfonts}%
\usepackage{amsthm}%
\usepackage{mathrsfs}%
\usepackage[title]{appendix}%
\usepackage{xcolor}%
\usepackage{textcomp}%
\usepackage{manyfoot}%
\usepackage{booktabs}%
\usepackage{algorithm}%
\usepackage{algorithmicx}%
\usepackage{algpseudocode}%
\usepackage{listings}%
\usepackage{bm} 
\usepackage{color}
\usepackage{enumerate}
\usepackage{dcolumn}
\usepackage{hyperref}
\usepackage{lineno}
\labelformat{section}{Section #1} 
\labelformat{subsection}{Section #1} 
\labelformat{equation}{Eq.~(#1)} 
\labelformat{figure}{Fig.~#1} 
\labelformat{subfigure}{Fig.~\thefigure#1} 
\labelformat{table}{Tab.~#1} 

\theoremstyle{thmstyleone}%
\theoremstyle{thmstyletwo}%

\theoremstyle{thmstylethree}%

\begin{document}

\title[{ }]{Distinguishing cosmological models through quantum signatures of primordial perturbations}

\author*[1]{Rathul Nath Raveendran}\email{rathulnath.r@gmail.com}

\author[1]{Sumanta Chakraborty}\email{tpsc@iacs.res.in}
\equalcont{These authors contributed equally to this work.}

\affil*[1]{School of Physical Sciences, Indian Association for the Cultivation of Science, Kolkata~700032, India}

\abstract{We study the evolution of various measures of quantumness of the curvature perturbation by integrating out the inaccessible entropic fluctuations in the multi-field models of inflation. In particular, we discuss the following measures of quantumness, namely purity, entanglement entropy and quantum discord. The models being considered in this work are ones that produce large scale curvature power spectra similar to those produced by single-field models of inflation. More specifically, we consider different multi-field models which generate nearly scale invariant and oscillatory curvature power spectrum and compare their quantum signatures in the perturbations with the corresponding single-field models. We find that, even though different models of inflation may produce the same observable power spectrum on large scales, they have distinct quantum signatures arising from the perturbation modes. This may allow for a way to distinguish between different models of inflation based on their quantum signatures. Intriguingly, this result generalizes to bouncing scenarios as well.}

\keywords{Inflation,  Multi-field models of inflation, Quantum state of primordial perturbations}



\maketitle

\section{Introduction}\label{sec:intro}

Inflation, which refers to a period of accelerated expansion of the universe, is typically modeled using scalar fields. Many models have been proposed, both those that involve a single scalar field and those that involve multiple scalar fields. It is also mentioned that while single-field inflationary models are in agreement with current observations of the cosmic microwave background (CMB) anisotropies, many models in high energy physics, such as those based on string theory, predict the presence of multiple scalar fields in the early universe~\cite{Baumann:2014nda,Yamaguchi:2011kg,Linde:1990flp}. This suggests that while single-field inflationary models may provide a good explanation for current observations, there may be additional complexity in the early universe that these models do not account for. The presence of multiple scalar fields can lead to distinct predictions, such as non-gaussianities, which can be used to test these models~\cite{Lalak:2007vi, Langlois:2008sg, Peterson:2010np, Kawai:2014gqa, Gao:2015aba, Braglia:2020eai, Braglia:2020fms, Palma:2020ejf}.

Inflationary models that involve multiple scalar fields, in addition to the usual curvature perturbations, entropic perturbations also arise. It is also known that the main observable feature of inflation on CMB is through curvature perturbations, and that the entropic modes are not directly observable~\cite{Ade:2015lrj}. Therefore, it is important to study the theoretical and observational differences between multi-field models and single-field inflation models in order to understand the effects of these entropic perturbations and distinguish them from the curvature perturbations. This could help in understanding the early universe and the underlying physics of inflation.

According to the theory of inflation, the quantum fluctuations of the scalar fields during the early universe are responsible for the primordial perturbations. The idea is that these quantum fluctuations become amplified during inflation and eventually become classical perturbations, which then provide the seeds for the formation of large scale structure in the universe \cite{Albrecht:1992kf,Polarski:1995jg,Kiefer:1998qe,Kiefer:1998jk,Kiefer:2006je,Martin:2007bw,Kiefer:2008ku,Martin:2012pea,Martin:2012ua,Martin:2015qta,Rajeev:2021lqk,Rajeev:2021yyl}. Recently, there is a growing interest in studying the quantum information contained in the cosmological perturbations \cite{Martin2022,Lim:2014uea,Maldacena:2015bha, Micheli:2022tld}. The focus of this work is to study the evolution of quantum signatures associated with the curvature perturbations in multi-field models of inflation. The study specifically focuses on quantum measures, such as entanglement entropy, of curvature perturbations in multi-field models, which are expected to evolve differently compared to single-field models. The work sets up the formalism for general multi-field inflation, and then specifically focuses on two field models of inflation which have a turning in the background trajectory.

Additionally, for multi-field models, if entropic fields are much heavier than the inflationary scale, one can compute a low energy theory for the adiabatic mode. This simplifies to an effective single-field theory, where the information about the original fields is encoded in the modified sound speed of adiabatic perturbations. Furthermore, it is known that the heavy (relative to the scale of inflation) degrees of freedom present in the model can generate the features in the curvature power spectrum~\cite{Achucarro:2010da,Chen:2014joa, Gao:2015aba}.  
Interestingly, some features generated in these models were found to enhance the fit to CMB data~\cite{Ade:2015lrj}.
If present, these features could lead to new tests of the multi-field paradigm and provide strong evidence for the existence of additional degrees of freedom during inflation. One of the main goals of this work is to examine the differences in quantum signatures between the original multi-field and effective single-field models.

Although inflation has been widely accepted, other explanations for the beginning of primordial perturbations have been studied. One frequently examined alternative is the ekpyrotic bouncing scenario \cite{Khoury:2001wf,Lehners:2007ac}. The ekpyrotic model offers the benefit of suppressing anisotropic instabilities that may occur during the contraction phase, due to the dominant energy density of the ekpyrotic source. However, single-field ekpyrotic models produce a strong blue tilt in the curvature perturbation spectrum \cite{Khoury:2001zk,Li:2013hga}. To address this, multiple-field models are proposed, where the contraction phase is dominated by isocurvature perturbations with a nearly scale-invariant spectrum. The additional field converts the isocurvature perturbations into adiabatic perturbations, resulting in a curvature perturbation spectrum that is consistent with observations \cite{Li:2013hga,Ijjas:2014fja}. One of our goals is to utilize our methods to analyze the evolution of quantum signatures linked to the curvature perturbations during the conversion process in the ekpyrotic models.

Our paper is organised as follows: In \ref{multifield}, we discuss the basic formalism regarding evolution of the quantum perturbations and the measure of quantumness in the context of multi-field models of early universe cosmology. Then we apply the above formalism in the context of two-field models of inflation in \ref{twofield} and discuss how the quantum measures evolve with the e-folding in two scenario --- (a) models with nearly scale invariant power spectrum and (b) models with oscillatory power spectrum. We also apply our methods to the ekpyrotic bouncing model of cosmology in \ref{ekpyrosis}. Finally, we comment on our results and provide possible future directions.

Notations and conventions: We shall work with the mostly positive signature convention, such that the Minkowski metric in the Cartesian coordinate becomes, ${\rm diag}.(-1,+1,+1,+1)$. Moreover, in the present work we will consider the background spacetime to be the spatially flat, Friedmann-Lema\^itre-Robertson-Walker (FLRW) metric, that is described by the line element,
\begin{equation}
\mathrm{d} s^2 = -\mathrm{d} t^2 + a^2(t)\,\delta_{ij}\, \mathrm{d} x^i\,\mathrm{d} x^j
=a^2(\eta)\, \left(-\mathrm{d}\eta^2+\delta_{ij}\, \mathrm{d} x^i\,\mathrm{d} x^j\right)~,
\label{eq:flrw-le}
\end{equation}
where $t$ and $\eta$ denote the cosmological time and the conformal time coordinates, respectively. Moreover, $a(t)$ represents the scale factor. We shall also adopt natural units wherein $\hbar=c=1$, and set the Planck mass to be $M_{_{\mathrm{Pl}}}=(8\, \pi\, G)^{-1/2}$. 

\section{Evolution of perturbations in the early universe with multi-field models}\label{multifield}

In this section, we study the evolution of the perturbations in the early universe cosmology, whose matter sector involves multiple scalar fields, driving the evolution of the universe. In particular, we assume that our model consists of $N_{\rm f}$ number of scalar fields. Therefore, when perturbed, the system will have $N_{\rm f}$ perturbing degrees of freedom arising from the $N_{\rm f}$ number of scalar fields and four scalar perturbing degrees of freedom from the metric. Intriguingly, all the $N_{\rm f}+4$ scalar perturbations are not necessary, due to gauge redundancies, and it is possible to eliminate four degrees of freedom in favour of the Mukhanov-Sasaki variables \cite{Mukhanov:1990me}. Among these, one will depend on the metric perturbations and is known as the comoving curvature perturbation, $v_\sigma$, and rest of the $(N_{\rm f}-1)$ quantities are referred to as the entropic perturbations. Moreover, we find it convenient to work with the phase space from the start. Since all these scalar perturbation modes with different wave numbers $k$ are decoupled, for convenience of the notation, we will not include the wave numbers explicitly in our analysis. In addition, we will be using the Schrödinger picture to perform quantization. In the Schrödinger picture, it is commonly recognized that working with real variables is a more convenient approach. Therefore, we will be using the real and imaginary parts of the Mukhanov-Sasaki variables for quantization purposes. By doing so, the Hamiltonian can be separated into two equivalent terms, which represent the independent evolution of the real and imaginary components (in this context, see Refs. \cite{Martin:2007bw, Martin:2012pea, Prokopec:2006fc, Battarra:2013cha, Raveendran:2022dtb}). With all these ingredients combined, the Hamiltonian for the real or imaginary parts of curvature and the entropy perturbations in the Fourier space can be written as,
\begin{equation} \label{eq:H}
H= \sum_{m,n}^{N_{\rm f}} \left(\frac{1}{2}p_{m}\,\delta_{mn}\,p_{n}
+p_m\,A_{mn}\,v_n+v_m\,B_{mn}\,v_n\right)~,
\end{equation}
where, $v_m$ are the real/imaginary parts of Mukhanov-Sasaki variables associated with the curvature and entropic perturbations and $p_m$ are the conjugate momenta. Also, $A_{mn}$ and $B_{mn}$ are both $(N_{\rm f}\times N_{\rm f})$ matrices, with entries being functions of the conformal time $\eta$. In particular, the matrix $B_{mn}$ is symmetric. These matrices denote the couplings between the momenta and the Mukhanov-Sasaki variables, as well as the coupling between the Mukhanov-Sasaki variables themselves, respectively. Note that, each of these variables and their conjugate momenta are functions of the wave number $k$ and should be expressed as $v_{m(k)}$ and $p_{m(k)}$, respectively, and the total Hamiltonian should also involve an integral over all the wave modes $k$. Since each of these wave modes are decoupled, it follows that, the results obtained for one of the modes will hold for all the others and hence we have removed any reference to the wave number from our analysis. 

From \ref{eq:H}, it is straightforward to obtain the Hamilton's equations as,
\begin{subequations}\label{eq:H-eqns}
\begin{align}
v_m'&=\frac{\partial H}{\partial p_{m}}=p_m + \sum_{n}^{N_{\rm f}} A_{mn} v_n\,,
\\
p_m'&=-\frac{\partial H}{\partial v_{m}}= -\sum_{n}^{N_{\rm f}} \left( p_n A_{nm} + 2 B_{mn} v_n\right)\,,
\end{align}
\end{subequations}
where, `prime' denotes derivative with respect to the conformal time $\eta$. Given the Hamiltonian, one can follow the canonical route to consider the quantum nature of these perturbations and hence one can construct the necessary expectation values, to construct the covariance matrix, in terms of these matrices $A_{mn}$ and $B_{mn}$ along with characteristics of the quantum state under consideration. This will enable us to determine various quantum measures, and as we will demonstrate these can be used as a tool to distinguish between single-field and multi-field models of inflation. We start by qunatizing the perturbations in the next section.  
\subsection{Quantizing the perturbations}

Quantizing the scalar perturbations in the cosmological background is equivalent to a time-dependent quantum mechanical problem in $N_{\rm f}$ dimensions. Since the Hamiltonian depends on the perturbation variables $v_{m}$, their conjugate momentum $p_{m}$ and also on the conformal time $\eta$, we can write down a Schr\"{o}dinger equation for the same, which simply reads, $i(\partial \Psi/\partial \eta)=H\Psi$. Here, the wave function $\Psi$ should be considered as a function of the Mukhanov-Sasaki variables $v_{m}$ and of conformal time $\eta$, i.e., $\Psi=\Psi(v_{m},\eta)$. Further, the momentum can simply be expressed as, $p_{m}=-i(\partial/\partial v_{m})$. Thus, in the Schr\"{o}dinger picture the evolution of the quantum state $\Psi(v_{m},\eta)$, using the Hamiltonian in \ref{eq:H}, is obtained from the following equation,
\begin{eqnarray}
i\,\frac{\partial \Psi}{\partial \eta} 
&=&\,\sum_{m,n}^{N_{\rm f}} \biggl[-\frac{1}{2}\delta_{mn}\left(\frac{\partial^2\Psi}{\partial v_m \partial v_{n}}\right) 
-i\, A_{mn}\,\left(\frac{1}{2}\Psi \, \delta_{mn}+\, v_n\, \frac{\partial \Psi}{\partial v_m}\right)  \nonumber \\
&+&\, B_{mn}\, v_m\,v_n\,\Psi\biggr]~.
\label{eq:se}
\end{eqnarray}
Note that this is the Schr\"{o}dinger equation corresponding to the quantum evolution of the Mukhanov-Sasaki variables. We would like to emphasize that the transition to the quantum domain with the above Hamiltonian has operator ordering ambiguity and as a simple choice, we have expressed the coupling term between the momentum and the Mukhanov-Sasaki variable as $(1/2)A_{mn}(p_{m}v_{n}+v_{n}p_{m})$ and that is reflected by the coefficient of the $A_{mn}$ term in the above Schr\"{o}dinger equation.

In order to proceed further, we need to come up with a suitable ansatz for the wave function $\Psi(v_{m},\eta)$. Since these perturbations are essentially vacuum fluctuations over the background solutions of the Einstein-Scalar field equations, we may consider these fluctuations to be in the ground state. Since the Hamiltonian is quadratic, the ground state can be taken to be Gaussian. Therefore, we may start with the following ansatz for the wave function,
\begin{equation}
\Psi(v_{m},\eta)=\mathcal{N}(\eta)\;
\mathrm{exp} \left[-\frac{1}{2}\,\sum_{r,s}^{N_{\rm f}}\,\Omega_{rs}(\eta)\, v_{r} \, v_{s}\right]~,
\label{eq:ga}
\end{equation}
where, $\Omega_{rs}$ is another symmetric $(N_{\rm f}\times N_{\rm f})$ matrix, with entries being functions of the conformal time $\eta$ and $\mathcal{N}(\eta)$ being the overall normalization factor. On substituting the Gaussian ansatz from \ref{eq:ga} in the Schr\"{o}dinger equation, presented in \ref{eq:se}, we find that the entries of the matrix $\Omega_{rs}$ satisfy the following differential equation,
\begin{equation}\label{eq:de-Omega}
i\, \Omega_{r s}'
=\sum_j^{N_{\rm f}}\Big(\Omega_{r j} \,\Omega_{j s}- i\, \Omega_{r j}\,A_{js}- i\, \Omega_{s j}\,A_{jr}-2 B_{rs}\Big)~.
\end{equation}
It is useful to write down the above expression in the matrix form, which reads,
\begin{equation}\label{eq:de-Omega-matrix}
i\,\boldsymbol{\Omega}'=\boldsymbol{\Omega}^{2}
-i\,\boldsymbol{\Omega}\,\boldsymbol{A}- i\, \boldsymbol{A}^{\rm T}\,\boldsymbol{\Omega}-2\boldsymbol{B}~.
\end{equation}
We will use this result in the later parts of this paper. The normalization of the wave function also depends on the matrix $\Omega_{rs}$, and yields the following relation between the functions~$\mathcal{N}(\eta)$ and~$\boldsymbol{\Omega}(\eta)$ to be,
\begin{equation}\label{eq:N-or}
\vert \mathcal{N}(\eta)\vert=\left(\frac{\mathrm{det.}\boldsymbol{\Omega}^{\mathrm{R}}}{\pi^{N_{\rm f}}}\right)^{1/4}~,
\end{equation}
where the superscript `R' to any quantity denotes its real part and the superscript `I' denotes its imaginary part. Thus we have the properties of the wave function, characterized by the matrix $\Omega_{mn}$ and the normalization factor $\mathcal{N}(\eta)$, determined from the above relations. 

Given the above wave function, one can compute the correlation functions between the Mukhanov-Sasaki variables and their conjugate momentum from \ref{eq:ga}, and these can be obtained as,
\begin{subequations}\label{eq:correlations}
		\begin{align}
			\left \langle  \hat{v}_{m}
			\hat{v}_{n}\right \rangle  
			=&\frac{1}{2}\left(\boldsymbol{\Omega}^{\mathrm{R}} \right)^{-1}_{mn}\equiv \alpha_{mn}~,\\
			\left \langle  \hat{p}_{m}
			\hat{p}_{n}\right \rangle =&\frac{1}{2}\left( \Omega^{\mathrm{R}}_{mn} + \sum_{i j}^{N_{\rm f}} \Omega^{\mathrm{I}}_{m i}\,(\boldsymbol{\Omega}^{\mathrm{R}})^{-1}_{ij} \Omega^{\mathrm{I}}_{jn}\right)\equiv \beta_{mn}~,\\
			\left \langle  \hat{v}_{m}
			\hat{p}_{n}+\hat{p}_{n}
			\hat{v}_{m}\right \rangle =&- \,\sum_j^{N_{\rm f}} (\boldsymbol{\Omega}^{\mathrm{R}})^{-1}_{mj} \, \Omega^{\mathrm{I}}_{jn} \equiv \gamma_{mn}~.
		\end{align}
\end{subequations}
For notational convenience we have introduced the matrices $\alpha_{mn}$, $\beta_{mn}$ and $\gamma_{mn}$, respectively. Further, note that the above relations presented in \ref{eq:correlations} suggests the following identity,
\begin{equation} \label{eq:invariant-multi}
4 \, {\bm \alpha} \, {\bm \beta}-{\bm \gamma}^2=\bm{I}~,
\end{equation}
where, $I_{mn}=\delta_{mn}$ is the identity matrix. Interestingly, none of the correlation functions depend on the details of the Hamiltonian explicitly, rather they depend on the wave function. However, there is an implicit dependence on the matrices $A_{mn}$ and $B_{mn}$ through the evolution equation for $\Omega_{mn}$, presented in \ref{eq:de-Omega-matrix}. Using this evolution equation for $\Omega_{mn}$ along with the definitions of the matrices $\alpha_{mn}$, $\beta_{mn}$ and $\gamma_{mn}$ from \ref{eq:correlations}, one can obtain the evolution equation for the two-point correlation functions as,
\begin{subequations}\label{eq:de-correlations}
		\begin{align}
			{\bm \alpha}'
			=&\bm{A}\,{\bm \alpha}+ {\bm \alpha} \, \bm{A}^{\rm T}+ \frac{1}{2}\left( {\bm \gamma} +{\bm\gamma}^{\rm T} \right)~,\\
 {\bm \beta}'=  &-\left[ {\bm \beta} \, \bm{A}+\bm{A}^{\rm T}\,{\bm \beta}+ \left(\bm{B}\,{\bm \gamma} +{\bm \gamma}^{\rm T}\, \bm{B}\right) \right]~,\\
 {\bm \gamma}'=& 2\,{\bm \beta} +\bm{A}\,\bm{\gamma} -\bm{\gamma} \, \bm{A}-4\,{\bm \alpha} \, \bm{B}~.
		\end{align}
\end{subequations}
By using the above equations, the evolution of the correlations between the Mukhanov-Sasaki variables and their conjugate momenta can be determined in the context of multi-field models, once the matrices $\bm{A}$ and $\bm{B}$ are fixed. One can also verify from the above evolution equations that the combination presented in \ref{eq:invariant-multi} indeed remains constant over time. Having discussed the basic evolution equations and the associated Schr\"{o}dinger equation, we now consider the measures of the quantumness of these perturbations.  

\subsection{Reduced Wigner function}

The first step in measuring the ``quantumness" of these perturbations is to construct the Wigner function, which is a tool used in the quantum mechanics to represent the state of a quantum system in phase space. It is a quasi-probability distribution because it can take on negative values, unlike a true probability distribution~\cite{Hillery:1983ms,case2008wigner}. 

Even though the Wigner function is not positive definite for all states of a quantum system, for the Gaussian wave function of our interest here, it turns out to be positive, and can be expressed as~\cite{case2008wigner, Martin2005, Raveendran:2022dtb},
\begin{align}
W(v_{m},p_{n})= \frac{1}{2\pi\sqrt{\textrm{det}~\bm{V}}}\textrm{Exp}\left[\frac{-\bm{Z}^T \bm{V}^{-1}\bm{Z}}{2}\right]~,
\label{Wig_exp_cov}
\end{align}
where, the vector $\bm{Z}$ is constructed out of the Mukhanov-Sasaki variables and their conjugate momentum, and the covariance matrix $\bm{V}$ depends on the matrices $\alpha_{mn}$, $\beta_{mn}$ and $\gamma_{mn}$, respectively. 
In general, the Wigner function will depend on the curvature perturbation, as well as the entropic perturbations. Since our interest lies in the curvature perturbations, we will focus only on the reduced Wigner function constructed out of the curvature perturbations alone. Therefore, the reduced vector $Z$ reads 
\begin{align}
Z=\begin{bmatrix}
v_\sigma \\
p_\sigma 
\end{bmatrix}~,
\label{def_Y_phase}
\end{align}
and the reduced covariance matrix $V$ becomes,
\begin{align}\label{covmat}
V&=\begin{bmatrix}
\alpha_{\sigma \sigma}  & \frac{1}{2}\gamma_{\sigma \sigma}
\\
\frac{1}{2}\gamma_{\sigma \sigma} & \beta_{\sigma \sigma}
\end{bmatrix}~.
\end{align}	
To be precise, the above correlations in general will also depend on the wave vector $\bm{k}$, such that we have the following relation: $\langle\hat{v}_{\bm k}\,\hat{v}_{\bm p}\rangle= \vert f_{k}\vert^2\,\delta^{(3)}({\bm k}-{\bm p})$, see Refs.~\cite{Martin:2012pea,Martin:2015qta}. Since the wave modes decouple among themselves, for convenience, we have dropped these delta functions. 

Given the above covariance matrix, the behavior of the Wigner function in the phase space can be tracked by following the evolution of the Wigner ellipse, as defined by the following relation,
\begin{equation}
\frac{\bm{Z}^T \bm{V}^{-1}\bm{Z}}{2}=1~.
\label{eq:we-tfm}
\end{equation}
It is important to note that area of the Wigner ellipse is proportional to $\textrm{det}.\bm{V}$ and it does not remain constant in multi-field models, in contrast to single field models, which will be a key distinguishing factor between multi-field versus single-field models of inflation. 
\subsection{Covariance matrix and its connection to quantum measures}

As mentioned earlier, our goal is to study the quantum measures associated with the adiabatic/curvature sector, which we refer to as our system, once the entropic degrees of freedom, namely the environment have been integrated out. It turns out that all of these quantum measures of the system can be determined from the covariance matrix for a Gaussian state after the entropic degrees of freedom have been traced over. This is because the covariance matrix contains all the observables, including those of the curvature sector and fully specifies its quantum state. Despite the fact that all of the quantum measures arise from the covariance matrix, it must be noted that each of these measures tell us about distinct properties of the system. Thus, by computing these measures one can infer about various quantum properties of the initial state as they evolve with the universe. For example, one of the important quantum property of the covariance matrix, that can be studied in detail, is known as decoherence, by which a pure quantum state becomes a statistical mixture due to its interaction with the environment. 
This can be measured by the so-called purity parameter and can be defined as~\cite{Serafini:2003ke, Walschaers:2021zvx}
\begin{equation}
p\equiv {\rm Tr}\left(\hat{\rho}^2_{\rm red}\right)~,
\end{equation}
where $\hat{\rho}_{\rm red}$ is the reduced density matrix associated with the quantum state of the co-moving curvature perturbation, when the entropic degrees of freedom are integrated out. This quantity determines whether the state is pure $\left(p =1\right)$ or, mixed $\left(p < 1\right)$. It is known that, for a Gaussian Wigner function, the purity is directly related to the determinant of the covariance matrix as \cite{Paris_2003,Adesso_2014},
\begin{equation}\label{purity}
p=\frac{1}{4\,\textrm{det.}\bm{V}}~.
\end{equation}
For single-field inflation, there are no entropic degrees of freedom and hence no environment. Therefore, no decoherence is present and hence the purity becomes unity. 
It is important to note, however, that higher-order non-Gaussian interaction terms in the Hamiltonian can introduce decoherence even in single-field models (in this context, see Ref.~\cite{Nelson:2016kjm,Gong:2019yyz,Burgess:2022nwu}). Multi-field inflationary models, on the other hand, have additional entropic degrees of freedom and hence the purity is less than unity. Thus, purity directly illustrates whether a state is pure/mixed, as well as distinguishes between single versus multi-field models. For multi-field models, entanglement entropy is another quantum measure, and arises from the entanglement between the additional entropic degrees of freedom and can be used to infer existence of multiple fields in the early universe. A commonly used measure of entanglement in a bipartite system is the von Neumann entropy $\mathcal{S}$ of either subsystem, as they are equal \cite{Prokopec:1992ia, Prokopec:2006fc,Battarra:2013cha} and defined as \cite{Adesso:2007tx,horodecki2009quantum},
\begin{equation}
\mathcal{S}=-\textrm{Tr}(\hat{\rho}_{\rm red}\ln \hat{\rho}_{\rm red})~,
\end{equation}
where, $\hat{\rho}_{\rm red}$ corresponds to the reduced density matrix, obtained by tracing over the irrelevant degrees of freedom. In our context, we are interested in the entanglement entropy associated with the curvature perturbation when the isocurvature/entropic perturbations are traced out. Moreover, it is known that for a Gaussian state, the entanglement entropy can be determined from the determinant of covariance matrix as \cite{Serafini:2003ke, Raveendran:2022dtb,Katsinis:2023hqn,Boutivas:2023ksg}
\begin{equation}
\mathcal{S} = F(\sqrt{4\,\textrm{det}.}\bm{V})~,
\label{eq:EE}
\end{equation}
with the function $F(x)$ being given by,
\begin{equation}
F(x) = \left(\frac{x+1}{2}\right)\, \ln\left(\frac{x+1}{2}\right) 
-\left(\frac{x-1}{2}\right) \ln\left(\frac{x-1}{2}\right)~.
\label{def_f}
\end{equation}	

Besides the purity and the entanglement entropy, another important quantum measure is the discord. It is well established that in pure states, quantum discord is equivalent to entanglement entropy \cite{datta2008quantum}. Since in the present context, we are investigating the multi-field inflationary model described by a pure quantum state $\Psi(v_{m},\eta)$, it is sufficient to compute the entanglement entropy in order to determine the discord. Hence, in what follows we will mostly concentrate on the entanglement entropy as the quantum measure. However, at a few instants, to further bolster the validity of our claim, we will also demonstrate that purity satisfies the desired relation, as expected.

Since all of these quantum measures are dependent on the determinant of the covariance matrix, it follows that once we know the evolution of the determinant of the covariance matrix, we can obtain the evolution of quantum measures, such as purity, entanglement entropy, and quantum discord. The determinant of the covariance matrix, $\textrm{det}.\bm{V}$ can be obtained by substituting the expectation values obtained in \ref{eq:correlations} in \ref{covmat}. Though the evolution of the covariance matrix depends on the matrices $A_{mn}$ and $B_{mn}$, which may appear to have a complex structure, in reality, these are simple ones. This is because, in the context of inflation, only the first entropic perturbation couples linearly to the adiabatic curvature perturbation~\cite{Pinol:2020kvw}. Therefore, many of the components of the matrix $A_{mn}$ and $B_{mn}$ vanishes, implying,
\begin{equation}
A_{n \sigma}=A_{\sigma n}=B_{n \sigma} = 0~, \qquad \textrm{for} \qquad n\geq 3.
\end{equation}
Then, using \ref{eq:de-correlations} one can obtain the evolution equations for the two-point correlation functions associated with the curvature perturbations as,
\begin{subequations}\label{eq:de-correlations2}
\begin{eqnarray}
\alpha_{\sigma \sigma} '&=& 2\left(A_{\sigma \sigma} \, \alpha_{\sigma \sigma} +  A_{\sigma s}\, \alpha_{s \sigma}\right)+\gamma_{\sigma \sigma}~,
\\
\beta_{\sigma \sigma}' &=&-2\left(\beta_{\sigma \sigma} A_{\sigma \sigma}+ \beta_{\sigma s}\, A_{s\sigma} + B_{\sigma \sigma} \, \gamma_{\sigma \sigma}  + B_{\sigma s} \, \gamma_{s \sigma}\right)~, 
\\
\gamma_{\sigma \sigma}' &=& 2 \, \beta_{\sigma \sigma} + A_{\sigma s} \, \gamma_{s \sigma} - \gamma_{\sigma s}\, A_{s \sigma} - 4 \left( \alpha_{\sigma \sigma} \, B_{\sigma \sigma} + \alpha_{\sigma s} \, B_{s \sigma}\right)~.
\end{eqnarray}
\end{subequations}
The above equations lead to the following evolution for the determinant of the reduced covariance matrix,
\begin{eqnarray}\label{eq:de-det}
\left(\textrm{det}.\bm{V}\right)'&= & 2\, A_{\sigma s}\left(\beta_{\sigma \sigma}\, \alpha_{s \sigma}  - \frac{\gamma_{\sigma \sigma}\, \gamma_{s\sigma}}{4}\right) - 2\, A_{s \sigma} \left(\alpha_{\sigma \sigma}\, \beta_{\sigma s} -\frac{\gamma_{\sigma \sigma}\, \gamma_{\sigma s}}{4}\right) \nonumber 
\\
&&+ 2\, B_{\sigma s} \left( \gamma_{\sigma \sigma} \, \alpha_{\sigma s} - \alpha_{\sigma \sigma} \, \gamma_{s \sigma}\right)~.
\end{eqnarray}
It is evident from the above expression that, the determinant of the covariance matrix evolves with the conformal time when the off-diagonal elements of matrices $A_{mn}$ and $B_{mn}$ are non-zero. This happens only when there is a coupling between the curvature and entropic degrees of freedom, as is the case here. 

The above analysis has been kept completely general and is applicable in a variety of scenarios involving an arbitrary number of scalar fields. However, to demonstrate the usefulness of this approach, we need to specialize to a certain type of early universe cosmological scenario, and the most simplest one being the two-field models. Since the two-field models have one curvature and one entropic degrees of freedom, interacting among themselves, it provides the best example to study the evolution of the quantum measures with the evolution of the universe. It will also serve our main purpose, i.e., distinguishing various models of early universe cosmology using quantum measures.

Before we proceed further, it is important to mention that the quantum measures we have discussed so far depend on the choice of creation and annihilation operators (see \cite{Kiefer:2006je, Martin:2015qta, Agullo:2022ttg} in this context). For example, in single-field models, quantum entanglement is typically studied for observables linked to wavenumbers ${\bf k}$ and ${\bf - k}$ and it requires a choice of creation and annihilation operators. However, there is no unique selection for these operators. And, in fact, given any Gaussian state in an FLRW spacetime, there always exists a choice of creation and annihilation operators for which
the state contains no entanglement between the ${\bf k}$ and ${\bf -k}$ sectors. Consequently, the notion of quantum measures becomes ambiguous. In this work, we focus on the entanglement between curvature and isocurvature perturbations, rather than the entanglement between ${\bf k}$ and ${\bf -k}$ sectors, with fixed creation and annihilation operators for these variables (see \cite{Prokopec:2006fc,Battarra:2013cha,Raveendran:2022dtb} in this context). However, while it is evident that entanglement is influenced by the choice of variables, it remains uncertain whether a canonical transformation exists that yields variables with zero entanglement between curvature and isocurvature perturbations. We aim to address this issue in our future work.
\section{Evolution of perturbations in the two-field models}\label{twofield}

The analysis presented in the above section is a general one, involving an arbitrary number of scalar fields. However, the main features described above are encoded in the simplest scenario, namely for two-field models. Thus, in this section, we will specialize to the two-field model involving one curvature and one entropic perturbation. It is known that the Hamiltonian associated with the Fourier components of the curvature and the entropic perturbations can be constructed as \cite{Raveendran:2022dtb},
\begin{equation} 
H_2=\frac{1}{2} \left(p_\sigma^2 + p_s^2\right) +p_\sigma \,\frac{z'}{z} \, v_\sigma+p_\sigma \, \xi \, v_s+p_s \, \frac{a'}{a} \, v_s +\frac{k^2}{2}\, v_\sigma^2 + \frac{1}{2}\left(k^2  + \mu_{s}^2\,a^2+\xi^2\right) v_s^2~,
\label{eq:H-tfm}
\end{equation}
where, $v_{\sigma}$ and $v_{s}$ are the curvature and the entropic perturbations, respectively. The momenta conjugate to these perturbations are denoted by, $p_\sigma$ and $p_s$. Note that the quantities $z(\eta)$ and $\xi(\eta)$ are the couplings between $p_{\sigma}$ with $v_{\sigma}$ and $v_{s}$, respectively, while the coupling between $p_{s}$ and $v_{s}$ depend on the scale factor $a(\eta)$. Moreover, $k$ corresponds to the magnitude of the wave vector and $\mu_{s}$ is a time-dependent mass scale associated with the entropic perturbation $v_{s}$.  From the above Hamiltonian (the subscript `2' in the Hamiltonian is to remind us that it is for two-field model), these two conjugate momenta can be obtained in terms of the perturbations and their conformal-time derivative as, 
\begin{equation}
p_\sigma=v_{\sigma}'-\frac{z'}{z}\, v_{\sigma}-\xi\, v_s~,
\qquad 
p_s= v_{s}'- \frac{a'}{a}\, v_s~.
\end{equation}
It is important to note that, in order to analyze specific models, the time-varying functions $\xi(\eta)$, $\mu_s(\eta)$, $z'/z$, and the scale factor $a(\eta)$ need to be fixed. Besides the above relations between the conjugate momenta and the time-derivative of the perturbations, the remaining Hamilton's equations of motion take the following forms,
\begin{eqnarray}
\label{eq:cl-H-eqs-tfm}
p_{\sigma}'= - \frac{z'}{z}\, p_{\sigma}-k^2\, v_{\sigma}~,
\qquad
p_{s}'= - \frac{a'}{a}\, p_{s}-\xi\, p_{\sigma}
-\left(k^2+\mu_{s}^2\,a^2 +\xi^2\right)\, v_{s}~.
\end{eqnarray}
Having described the evolution equations for the perturbations and their conjugate momentum, let us concentrate on the entries of the matrices $A_{mn}$ and $B_{mn}$, responsible for the evolution of the quantum measures. These matrices can be read out from the Hamiltonian for the two-field model, as described by \ref{eq:H-tfm}, and one can identify them as,
\begin{align}
A&=\begin{bmatrix}
\frac{z'}{z}  & \xi  \\
0  & \frac{a'}{a}  
\end{bmatrix}~,
\quad B=\begin{bmatrix}
\frac{k^2}{2}  & 0  \\
0  & \frac{k^2 + a^2 \mu_s^2 +\xi^2}{2}  
\end{bmatrix}~.
\end{align}
As expected, the matrix $B$ is symmetric, but the matrix $A$ is not. Then from \ref{eq:de-Omega-matrix}, the evolution of the quantities $\Omega_{\sigma\sigma}$, $\Omega_{ss}$, and $\Omega_{\sigma s}$, present in the Gaussian wave function describing the state of the quantum perturbations in the two-field model, is described by the following differential equations,
\begin{subequations}\label{eq:de-Omega-tfm}
\begin{eqnarray}
i\, \Omega_{\sigma \sigma}'&=& \Omega_{\sigma\sigma}^2  + \Omega_{\sigma s}^2 - 2 \, i\, \frac{z'}{z} \, \Omega_{\sigma\sigma}- k^2~,
\\
i\, \Omega_{ss}'&=& \Omega_{ss}^2+ \Omega_{\sigma s}^2  - 2\,i\, \frac{a'}{a} \, \Omega_{ss}- 2\,i \,\xi\,\Omega_{\sigma s}
- \left(k^2 + a^2 \mu_s^2 +\xi^2\right)~,
\\
i\, \Omega_{\sigma s}'&=& \Omega_{\sigma \sigma}\,\Omega_{\sigma s}+\Omega_{\sigma s}\,\Omega_{ss}- i \left(\frac{z'}{z}+\frac{a'}{a}\right)\,
\Omega_{\sigma s} - i\, \xi\,\Omega_{\sigma \sigma}~.
\end{eqnarray}
\end{subequations}
The determinant of the covariance matrix, namely $\textrm{det}.\bm{V}$, can be obtained in terms of the entries of the Gaussian wave function as, 
\begin{equation}
\textrm{det}.\bm{V} =\frac{1}{4}\left(1+ \frac{\vert \Omega_{\sigma s} \vert^2}{\textrm{det.}\boldsymbol{\Omega}^{\mathrm{R}}}\right)~.
\end{equation}
Further, using \ref{eq:de-correlations} one can obtain the evolution equations for the two-point correlation functions associated with the curvature perturbations as,
\begin{subequations}\label{eq:de-correlations-sigma}
\begin{align}
\left \langle \hat{v}_{\sigma}^{2}\right \rangle '
=& 2\frac{z'}{z}\,\left \langle  \hat{v}_{\sigma}
			\hat{v}_{\sigma}\right \rangle + \left \langle  \hat{v}_{\sigma}
			\hat{p}_{\sigma}+\hat{p}_{\sigma}
			\hat{v}_{\sigma}\right \rangle+2\,\xi\,\left \langle  \hat{v}_{s}
			\hat{v}_{\sigma}\right \rangle~,
\\
\left \langle  \hat{p}_{\sigma}^{2}\right \rangle' 
=&k^2\, \left \langle  \hat{v}_{\sigma}
			\hat{p}_{\sigma}+\hat{p}_{\sigma}
			\hat{v}_{\sigma}\right \rangle-2\frac{z'}{z}\, \left \langle  \hat{p}_{\sigma}
			\hat{p}_{\sigma}\right \rangle~, 
\\
\left \langle  \hat{v}_{\sigma}\hat{p}_{\sigma}+\hat{p}_{\sigma}\hat{v}_{\sigma}\right \rangle' 
=& 2 \left \langle  \hat{p}_{\sigma}
			\hat{p}_{\sigma}\right \rangle - 2 k^2\, \left \langle  \hat{v}_{\sigma}
			\hat{v}_{\sigma}\right \rangle 
   +\xi\, \left \langle  \hat{v}_{s}
			\hat{p}_{\sigma}+\hat{p}_{\sigma}
			\hat{v}_{s}\right \rangle~.
		\end{align}
\end{subequations}
The above equations lead to the evolution equation for the determinant of the covariance matrix in the two-field model as,
\begin{equation}\label{eq:detV-de-2f}
  \left(\textrm{det}.\bm{V}\right)' = 2\,\xi\,C_{s \sigma}\,,
\end{equation}
where the quantity  $C_{s \sigma}$ is defined as
\begin{equation}\label{eq:Cssigma-def}
  C_{s \sigma}\equiv \left \langle  \hat{v}_{s}
			\hat{v}_{\sigma}\right \rangle \left \langle  \hat{p}_{\sigma}
			\hat{p}_{\sigma}\right \rangle -\left \langle  \hat{v}_{s}
			\hat{p}_{\sigma}\right \rangle \frac{\left \langle  \hat{v}_{\sigma}
			\hat{p}_{\sigma}+\hat{p}_{\sigma}
			\hat{v}_{\sigma}\right \rangle}{2}\,.
\end{equation}
Recall that, once we know the evolution of $\textrm{det}.\bm{V}$, we can obtain the evolution of the quantum measures, such as purity, entanglement entropy and quantum discord. At this stage, it is also useful to introduce the definition of power spectra associated with curvature, $P_{_\mathcal{R}}$ and isocurvature perturbation, $P_{_S}$ as
\begin{subequations}\label{eq:power-spectra-defs}
\begin{align}
P_{_\mathcal{R}}& \equiv\frac{k^3}{2\,\pi^2}\frac{\left \langle \hat{v}_{\sigma}^{2}\right \rangle}{z^2}~,\\
P_{_S}&\equiv \frac{k^3}{2\,\pi^2}\frac{\left \langle \hat{v}_{s}^{2}\right \rangle}{z^2}~.
		\end{align}
\end{subequations}
We will use these definitions of the power spectrum for the two perturbations in the later parts of this work.

Having discussed the evolution of relevant quantities, let us come back to the perturbation Hamiltonian $H_{2}$ for the two-field model. As evident from \ref{eq:H-tfm}, the curvature and entropic perturbations are disjoint, except for the coupling function $\xi(\eta)$. Whenever, $\xi$ vanishes, the curvature and the entropic perturbations evolve independently, while the presence of $\xi$ makes the evolution of these perturbations to be dependent on one another. The origin of this coupling can be traced to the interaction between the two background fields in the two-field model. Thus in the field space, the trajectory of these two background fields deviates from their free evolution due to the coupling and hence the onset of the coupling term is often referred to as the ``turning" in the background trajectory \cite{Lalak:2007vi,Achucarro:2010da}.  

Our next task is to apply the formalism for two-field model, presented above, to investigate the impact of turning in the background trajectory in the evolution of $\textrm{det}.\bm{V}$ in the context of two-field models. In other words, we wish to explore, how the coupling function $\xi(\eta)$ affects the evolution of the determinant of the covariance matrix. For this purpose, we consider two scenarios, first, we consider a model in which the turning of the background trajectory occurs well after the cosmologically relevant scales exit the Hubble horizon. In these models, the modes associated with the large scale curvature perturbations are not affected by the entropic perturbations, since the coupling $\xi(\eta)$ was turned off, till the curvature perturbations exit the Hubble horizon. In the field space, it results into no change in the trajectory of the fields. In another model, the turning is introduced during the Hubble exit of the relevant curvature perturbation scales. As a consequence, in contrast to the first case, the presence of non-zero coupling $\xi(\eta)$ in this model leads to features in the power spectrum over the curvature scales relevant for CMB. We will further demonstrate that the power spectra generated in these models, with distinct features, can also be generated in single-field models. In particular, we will explicitly demonstrate how the model with non-zero coupling function, is equivalent to a single-field model of inflation, using the effective field theory approach. 

\subsection{Models with nearly scale-invariant power spectrum}
\label{sec:model-SI}
First, we examine models that produce a nearly scale-invariant power spectrum, suggested by observations. As emphasized before, in this case, the coupling function sets in after the relevant curvature perturbation modes have exited the Hubble horizon. Therefore, the curvature perturbations responsible for the power spectrum observed today do not interact with the entropic perturbations at all, leading to scale invariant power spectrum. In order to see these explicitly, and also to study the evolution of the quantum measures, including the determinant of the covariance matrix, we need to specify the time evolution of the functions $z$, $\xi$ and $\mu_{s}$. To do this, we need to specify the background evolution, which we consider to be originating from an inflationary scenario. Thus, in the context of slow-roll inflation, we introduce the following slow-roll parameters, 
\begin{subequations}\label{eq:def-epsilons}
		\begin{align}
    \epsilon_1=& - \frac{{\rm d\, log}H}{{\rm d} N}~,
    \\
    \epsilon_{n+1}=& \frac{{\rm d\, log}\epsilon_n}{{\rm d} N}~,
		\end{align}
\end{subequations}
where, $H$ is the Hubble parameter and $N$ corresponds to the e-folding parameter, defined as, 
\begin{equation}
N(\eta)=\int_{\eta}^{\eta_{\rm end}}aH\,d\eta~.
\end{equation}
Here, $\eta_{\rm end}$ corresponds to the end of the inflation. Note that the e-folding parameter is a monotonic function of the conformal time in the inflationary epoch and hence can be used as a proxy for the clock. Thus all the time derivative operators can be replaced by derivatives with respect to the e-folding parameter $N$, as in \ref{eq:def-epsilons}. It is convenient to write $(z'/z)$, in terms of the second slow roll parameter $\epsilon_{2}$ as,
\begin{equation}\label{eq:zpz-in-es}
\frac{z'}{z}=a\, H\left( 1+\frac{\epsilon_2}{2}\right)~,
\end{equation}
where, as pointed out before $H$ is the Hubble parameter and the slow roll parameters are defined in \ref{eq:def-epsilons}. 

The simplest method to obtain models with nearly scale-invariant power spectrum is by selecting a constant second slow-roll parameter. Then the first slow-roll parameter and the Hubble parameter can be obtained using \ref{eq:def-epsilons} as,
\begin{subequations}\label{eq:e1-H}
\begin{eqnarray}
\epsilon_1(N)&=&\epsilon_1(N_p) \, {\rm exp}\left[\epsilon_2\left( N-N_p\right)\right]~,
\\
H(N)&=&H(N_p)\,{\rm exp}\left[\frac{\epsilon_1(N) - \epsilon_1(N_p)}{\epsilon_2}\right]~,
\end{eqnarray}
\end{subequations}
where $N_p$ is the time (expressed in terms of e-folding) at which the pivot scale $k=0.05\,{\rm Mpc^{-1}}$ exits the Hubble horizon. For the coupling function $\xi$, on the other hand, which introduces a change in the trajectory at a point well after the cosmologically relevant scales have exited the Hubble horizon, so that it does not affect the evolution of the curvature perturbations. Therefore, we select the coupling parameter $\xi$ in a simple form as proposed in \cite{Achucarro:2010da} for our purpose as,
\begin{equation}\label{eq:xibaH}
\frac{\xi}{a H} = \xi_{m} \, {\rm sech}^2\left[\frac{(N-N_0)}{\Delta}\right]~,
\end{equation}
where, $\xi_{m}$, $N_0$, and $\Delta$ are constants. Note that the e-folding parameter $N_{0}$ signifies the time, when the coupling between curvature and entropic perturbation is important. Similarly, the mass $\mu_s$ associated with the entropic perturbation can be expressed as,
\begin{equation}
\mu_s^2= M^2-\left(\frac{\xi}{2a}\right)^2=M^{2}\left[1-\left(\frac{H \xi_{m}}{2M}\right)^{2}{\rm sech}^4\left[\frac{(N-N_0)}{\Delta}\right] \right], 
\end{equation}
where, the quantity $M$ is another constant. Note that for large enough $\xi_{m}$, in particular, if $\xi_{m}>(2M/H)$, the mass of the entropic perturbation will be negative near $N=N_{0}$. This is acceptable as long as it does not result in instability (see for example \cite{Renaux-Petel:2015mga} for more details). It is also worthwhile to point out that, the fact that $\mu_s^2$ can take negative values makes it possible to generate primordial black holes in two-field inflationary models (see \cite{Palma:2020ejf} in this context). 

\begin{figure}[h!]
\centering
\includegraphics[width=0.75\linewidth]{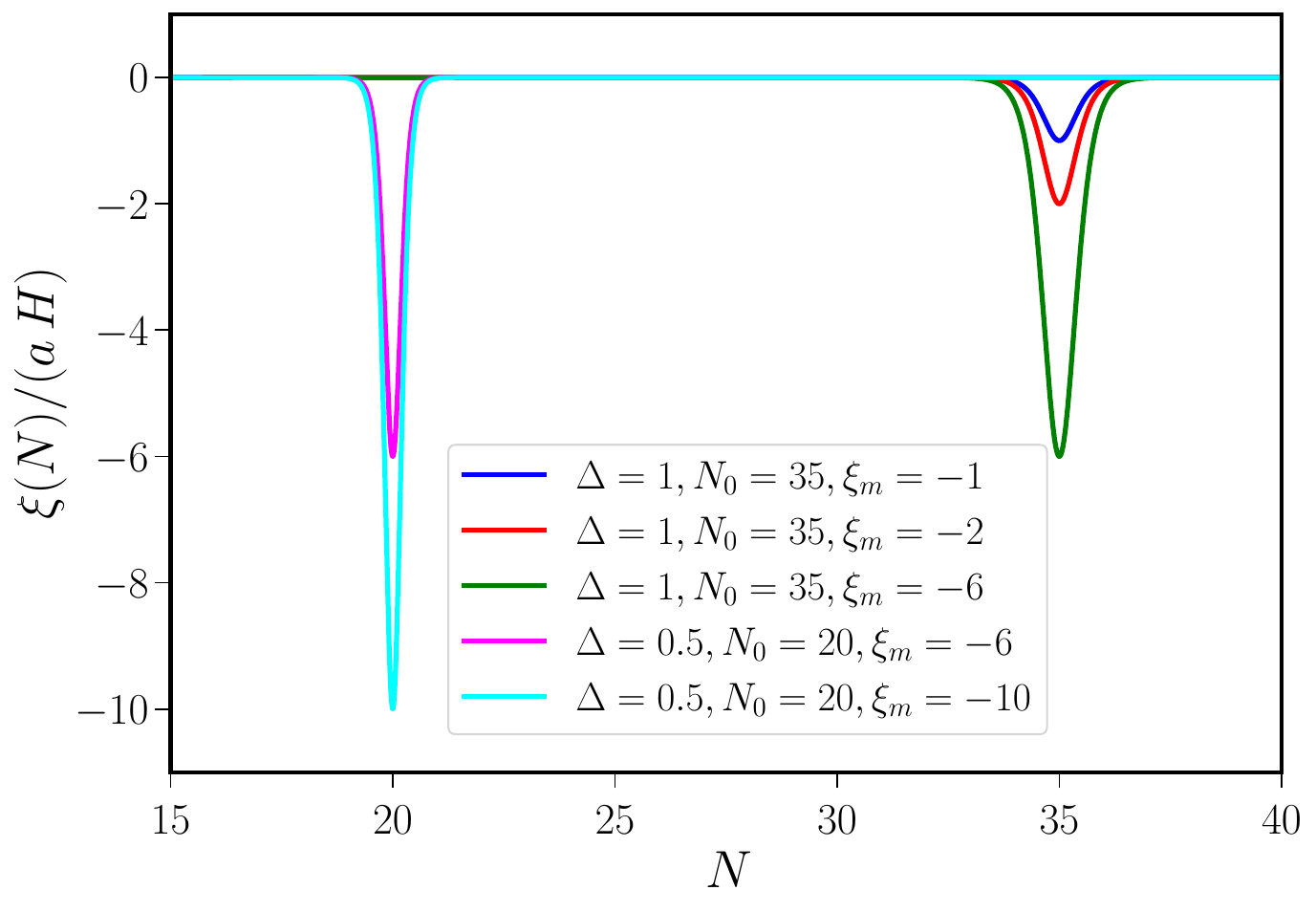}
\caption{The evolution of coupling parameter $\xi$ has been presented for different values of the amplitude $\xi_m$, the central location $N_{0}$ and the width $\Delta$. A non-zero value for $\xi$ signifies a turning in the background trajectory. It is important to note that while the $\xi$ peaking at $N_0=20$ leads to features in the curvature power spectrum, the $\xi$ with maximum contribution at $N_0=35$ does not effect the evolution of curvature power spectrum. Since the pivot scale exits the Hubble horizon at an e-folding parameter larger than 20, but smaller than 35. These plots are made with $\epsilon_2=0.036$, $\epsilon_1(N_p) = 1.875\times10^{-3}$, and $H(N_p) =1.724 \times 10^{-5} M_{_{\mathrm{Pl}}}$, which results into the scalar spectral index to be $n_{\rm s} =0.96$ and the amplitude of the scalar power spectrum as $ A_{\rm s} = 2\times 10^{-9}$. Further, for the curves with $N_{0}=35$, we choose to work with $M^2= H^2$ and for $N_0=20$, we choose $M^2=300H^2$.}
\label{fig:xibaH-N0-35}
\end{figure}

\begin{figure}[h!]
\centering
\includegraphics[width=0.45\linewidth]{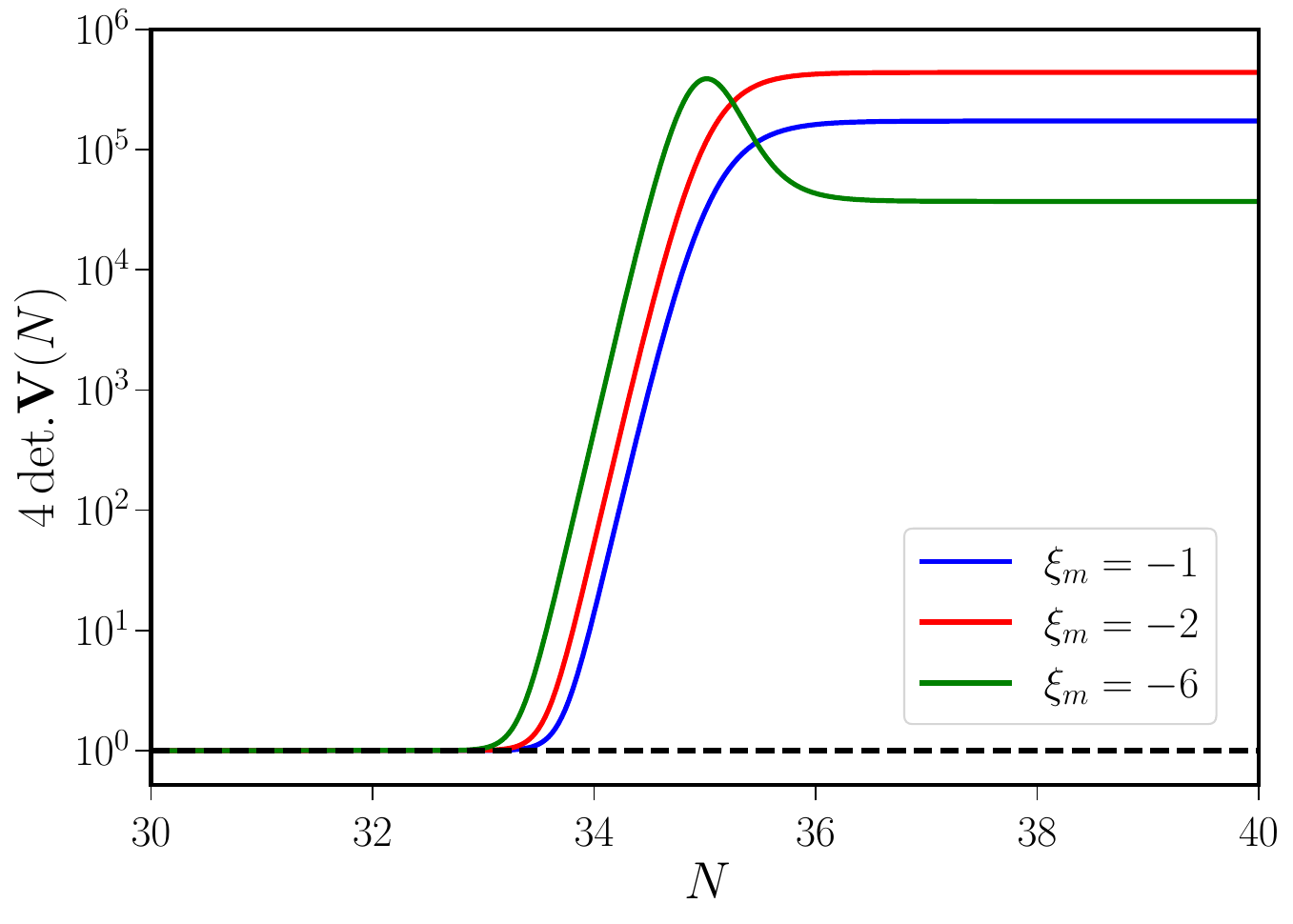} ~~
\includegraphics[width=0.45\linewidth]{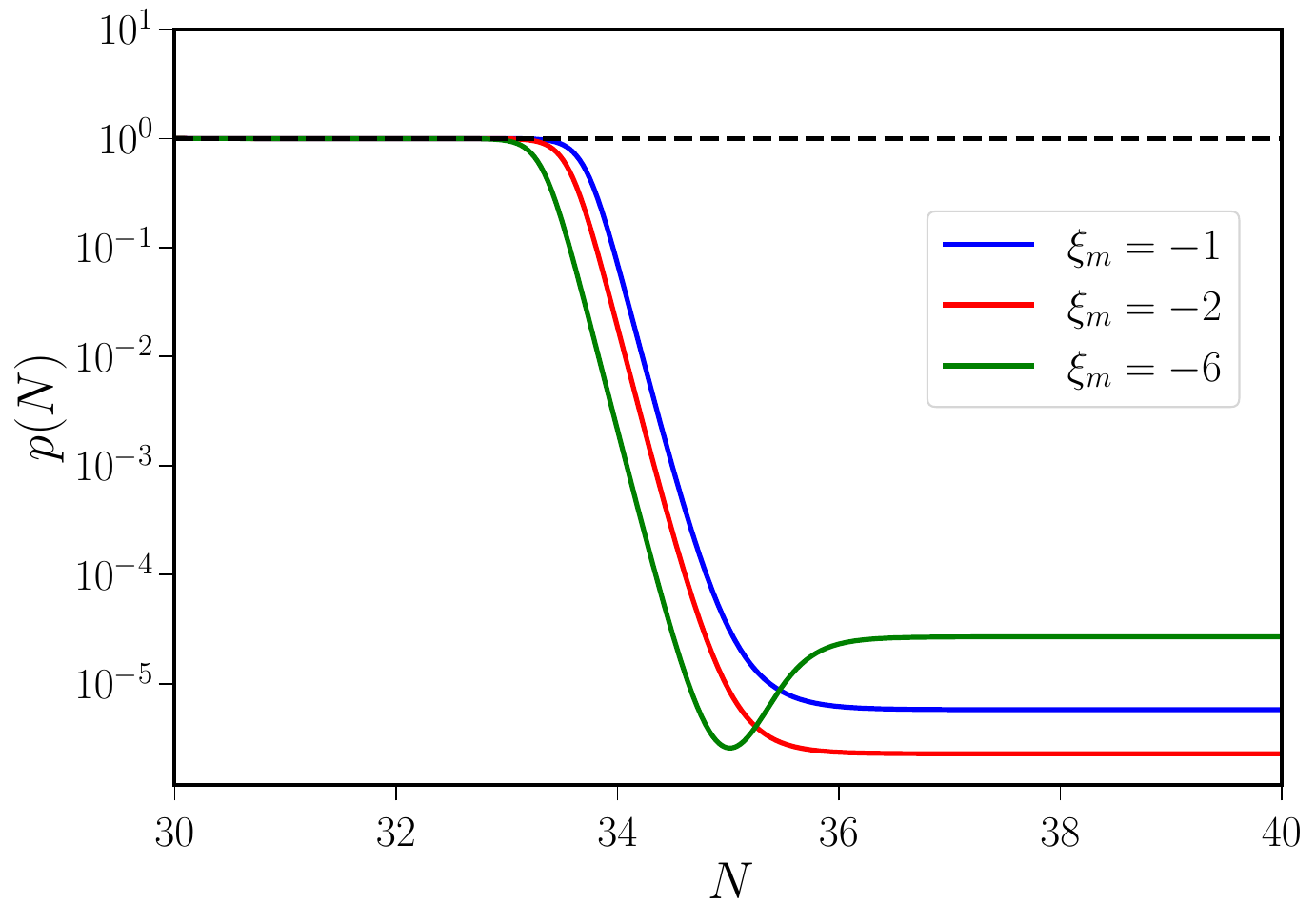}\\
\includegraphics[width=0.5\linewidth]{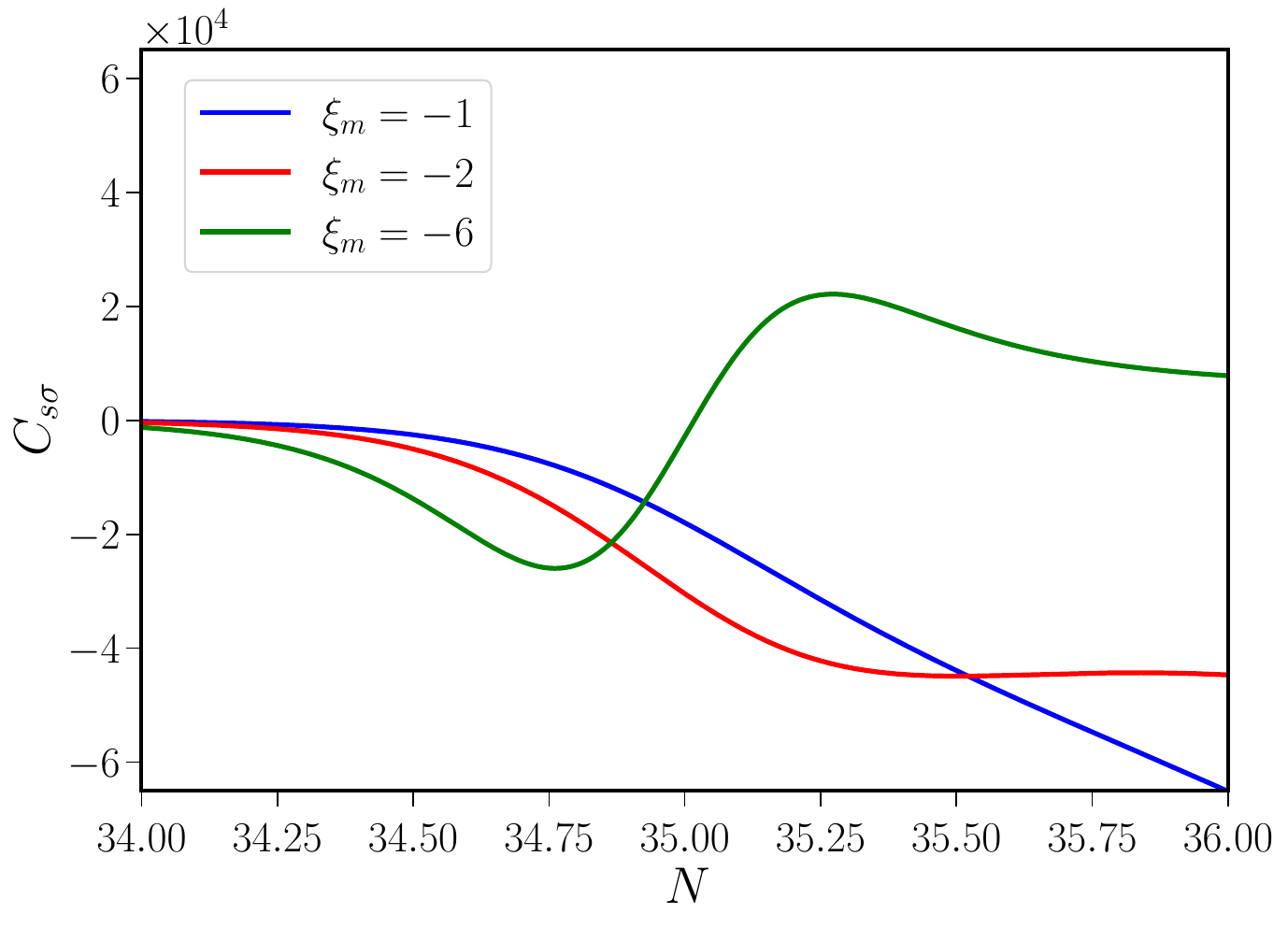}
\caption{Comparison of the effect of turning in the trajectory, depicting the evolution of ${\rm det}.V$, $p$ and $C_{s \sigma}$, when the coupling $\xi$ peaks at $N_0=35$, for different values of $\xi_m$. The evolution is associated with the pivot scale, located at $k=0.05\,{\rm Mpc^{-1}}$.}
\label{fig:detV-N-N0-35}
\end{figure}

To study the effect of the coupling parameter $\xi$, on the change in the trajectory on the evolution of $\textrm{det}.\bm{V}$, along with the modifications in the quantum measures, we consider $N_{0}>N_{\rm p}$, i.e., the interaction is switched on after the pivot scale has exited the Hubble radius. Following which, with reasonable choices of the parameters appearing in the problem, e.g., $\epsilon_{2}$, $N_{\rm p}$, we could get the scalar spectral index $n_{\rm s}$ and scalar power spectrum amplitude to be consistent with the Planck data and plotted the evolution of the quantity $\xi$ with the e-folding parameter in \ref{fig:xibaH-N0-35}. In this figure, we have also included the evolution of $\xi$, characterizing the strength of the interaction between the curvature and the isocurvature perturbations, for different values of $\xi_{m}$. Following this evolution of the coupling function $\xi$, in \ref{fig:detV-N-N0-35}, we have depicted how turning on the coupling the turning in field space causes a change in the evolution of $\textrm{det}.\bm{V}$, purity $p$ and $C_{s \sigma}$. As evident, once the coupling again gets turned off, $\textrm{det}.\bm{V}$ reaches a constant value, similar to what happens in the single field case. A similar behaviour is also seen for purity. However, the quantity $C_{s\sigma}$ keeps evolving non-trivially even after the coupling has been turned off and depends on the choice of $\xi_{m}$. 

The behavior of $\textrm{det}.\bm{V}$ can be explained by \ref{eq:detV-de-2f}. Since the coupling vanishes after certain duration of time (see \ref{fig:xibaH-N0-35}), the rate of change of the determinant of the covariance matrix also goes to zero, and hence $\textrm{det}.\bm{V}$ becomes constant in time. Moreover, for $\xi_{m}=-1$ and $\xi_{m} = -2$, the right hand side of \ref{eq:detV-de-2f} is always positive and hence $\textrm{det}.\bm{V}$ increases after the turn. On the other hand, when $\xi_{m} =-6$, the sign of $C_{s \sigma}$ changes during the turn, which causes a decrease in the evolution of $\textrm{det}.\bm{V}$. This in turn also fixes the behaviour of the purity $p$. As evident from \ref{purity}, $p$ will show an exactly opposite behaviour to $\textrm{det}.\bm{V}$. In particular, for $\xi_{m}=-1$ and $\xi_{m} = -2$, the purity decreases, as $\textrm{det}.\bm{V}$ after the turn. While for $\xi_{m} =-6$, the purity increases after the turn. This highlights the possibility of restoring purity in the system through judicious parameter selection, known as recoherence \cite{Colas:2022kfu}.

\begin{figure}[h!]
\centering
\includegraphics[width=0.45\linewidth]{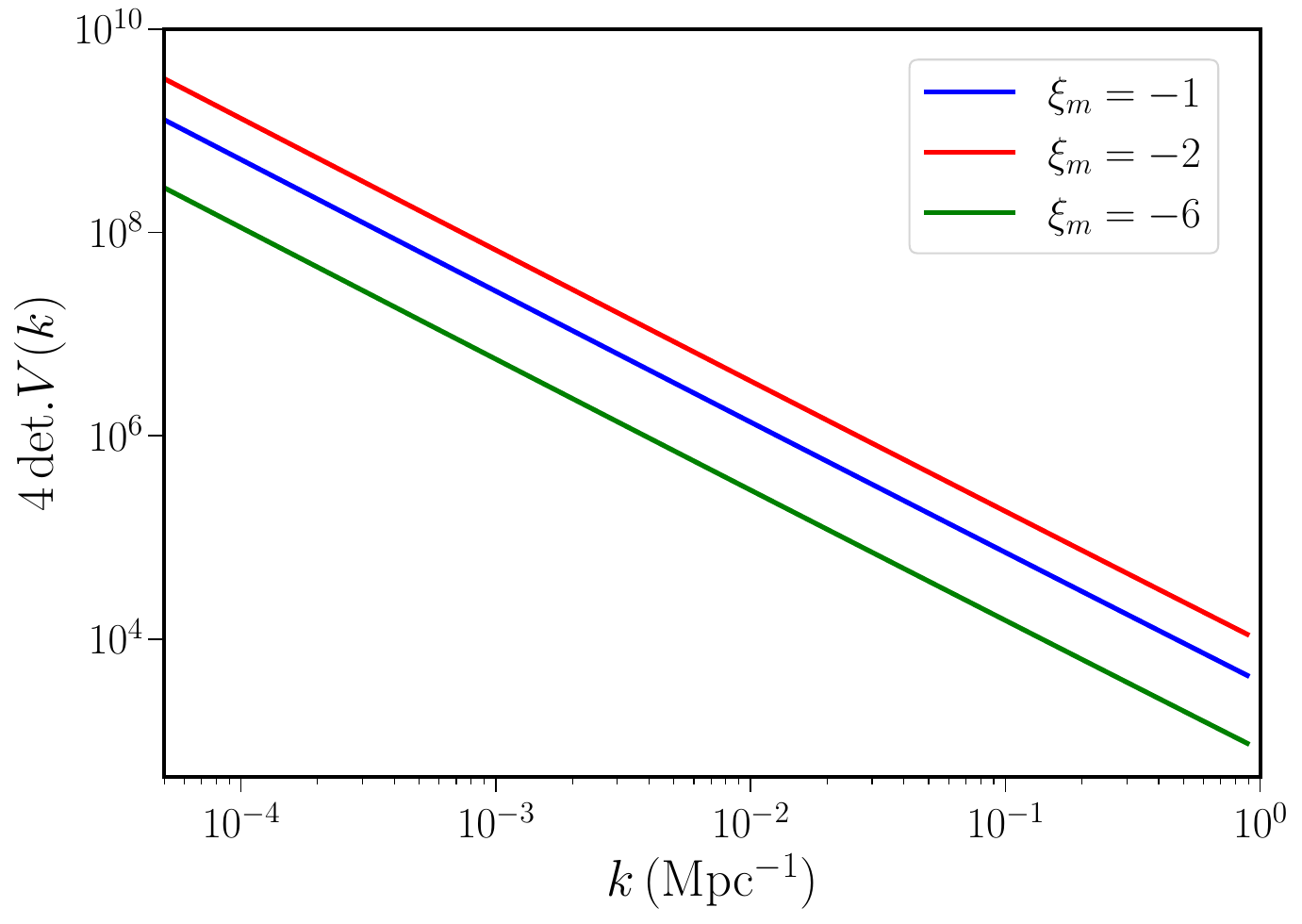}~~
\includegraphics[width=0.45\linewidth]{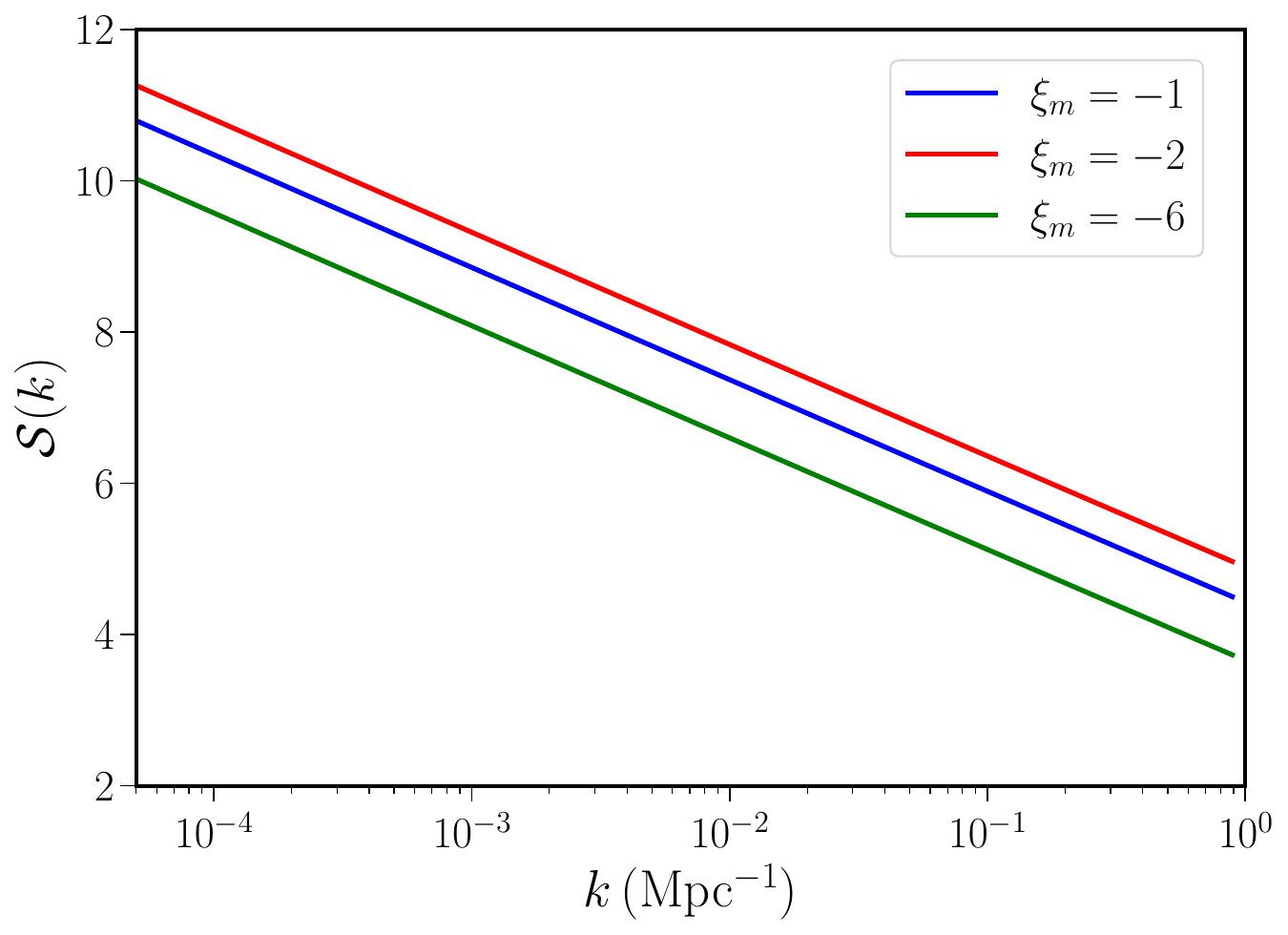}
\caption{The power spectra associated with the determinant of the covariance matrix, and the entanglement entropy have been presented, after the turning for different values of $\xi_m$. The amplitudes and the form of $\xi$ are chosen in such a way that it does not effect the power spectrum of curvature perturbations, which remains nearly scale invariant.}
\label{fig:deltak}
\end{figure}

Finally, in \ref{fig:deltak}, we present the power spectra of $\textrm{det}.\bm{V}$ and the entanglement entropy $\mathcal{S}$ for wave numbers relevant for the CMB observations. It can be observed that the spectra of $\textrm{det}.\bm{V}$ and the entanglement entropy $\mathcal{S}$ are enhanced compared to the single field models, and this is because a turn in the field space occurs, due to non-zero $\xi$. It is worth noting that in single-field models of inflation, the value of $\textrm{det}.\bm{V}$ is always equal to unity and hence is an important measure to distinguish multi-field models of inflation from a single-field one. It is worth re-emphasizing that the power spectrum of the curvature perturbation remains unchanged during the turn introduced in the field space since the turn occurs well after the observable modes of curvature perturbation exit the Hubble radius. 

\subsection{Models with features in the power spectrum}
\label{sec:model-features}
Our next task is to study the effect of the turn on the quantum state of the perturbations, when the turn occurs during the exit of cosmologically relevant scales through the Hubble horizon, leading to distinct features in the power spectrum of curvature perturbation. An interesting aspect of these models being, they can be matched to an equivalent single-field model producing the same power spectrum \cite{Achucarro:2010da}. However, these two-field and single-field models are distinguishable through the quantum measures, which have distinct signatures. We will show this result shortly. 

\begin{figure}[]
\centering
\includegraphics[width=.75\linewidth]{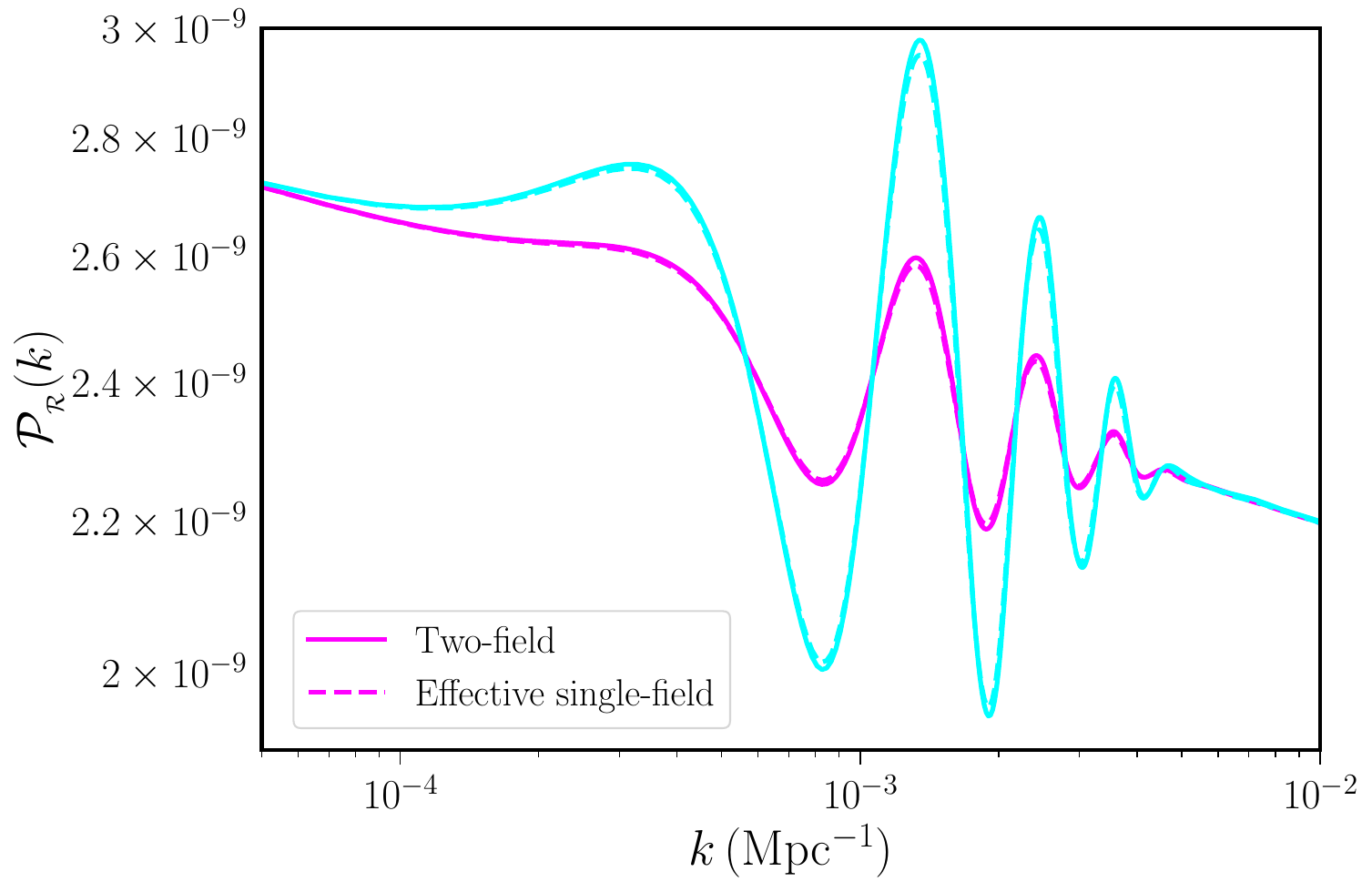}
\caption{The power spectra of curvature perturbations generated during the two-field model of inflation is compared with the power spectra arising from the effective single-field inflationary scenario. We have depicted two possible amplitudes of the coupling function, cases with $\xi_m =-6$ (represented in magenta) and $\xi=-10$ (represented in cyan). See the text for more discussions.}
\label{fig:PR-2f-1f-EFT}
\end{figure}

To start by showing the equivalence of the two-field model considered here with a single-field model of inflation. It is known that, in the particular case of multi-field models, if the entropic degrees of freedom are sufficiently massive compared to the scale of inflation, then it is possible to obtain an effective low energy theory for the adiabatic/curvature perturbation mode alone. In other words, one can obtain an effective single-field Hamiltonian from the general Hamiltonian in \ref{eq:H}, if entropic degrees of freedom are sufficiently massive. In what follows, we will concentrate on the two-field model of inflation considered in the present section. Since the entropic degree of freedom is massive, it can be considered as classical, and hence the curvature perturbation dependent part of the Hamiltonian in \ref{eq:H-tfm} can be written as,
\begin{equation} 
H_{\sigma}= \frac{\bar{p}_\sigma^2}{2} + \bar{v}_\sigma \frac{\bar{z}'}{\bar{z}}\bar{p}_\sigma +c_s^2 \, k^{2}\, \bar{v}_\sigma^2,\label{eq:H-1f-eft-general}
\end{equation}
where, 
\begin{equation}\label{eq:eqs-2f-bars}
c_s^2\equiv 1+\xi(\eta) \,\frac{v_s}{p_\sigma}, \quad
\bar{z}\equiv \frac{z}{c_s},\quad \bar{v}_\sigma\equiv \frac{v_\sigma}{c_s}, \quad
\bar{p}_\sigma\equiv c_s \, p_\sigma.
\end{equation}
Since $c_{s}$ depends on $p_{\sigma}$, the above cannot be considered as a true Hamiltonian. To circumvent the same, we need to find out a relation between $v_{s}$ and $p_{\sigma}$, depending on the background quantities. Since the entropic degree of freedom is assumed to be sufficiently massive, then one can indeed express the ratio $(v_{s}/p_{\sigma})$ in terms of background quantities as,
\begin{equation}
\frac{v_s}{p_\sigma}=\frac{\xi} {\left(\frac{a''}{a}-\mu_{s}^2\,a^2 -k^2-\xi^2\right)}~. 
\end{equation}
The above relation leads to the following expression,
\begin{equation}
c_s^2 = \left[1+ \frac{\xi^2} {\left(k^2- \frac{a''}{a}+\mu_{s}^2\,a^2\right)}\right]^{-1}~.
\end{equation}
Thus one can conclude that $H_{\sigma}$, presented in \ref{eq:H-1f-eft-general}, is indeed an effective Hamiltonian for the curvature perturbation, whose power spectrum would be similar to that of a single field model of inflation. For completeness, we have plotted the power spectrum arising from the two-field model of inflation and have compared the same with the power spectrum of the effective single-field model in \ref{fig:PR-2f-1f-EFT}. As evident, the power spectrum of the two-field model closely follows the power spectrum of the effective single-field scenario. However, the power spectrum depends on the amplitude $\xi_{\rm m}$ of the coupling function, as different $\xi_{\rm m}$ leads to very different values for the power spectrum, though the overall structure remains the same. 

\begin{figure}[h!]
\centering
\includegraphics[width=0.45\linewidth]{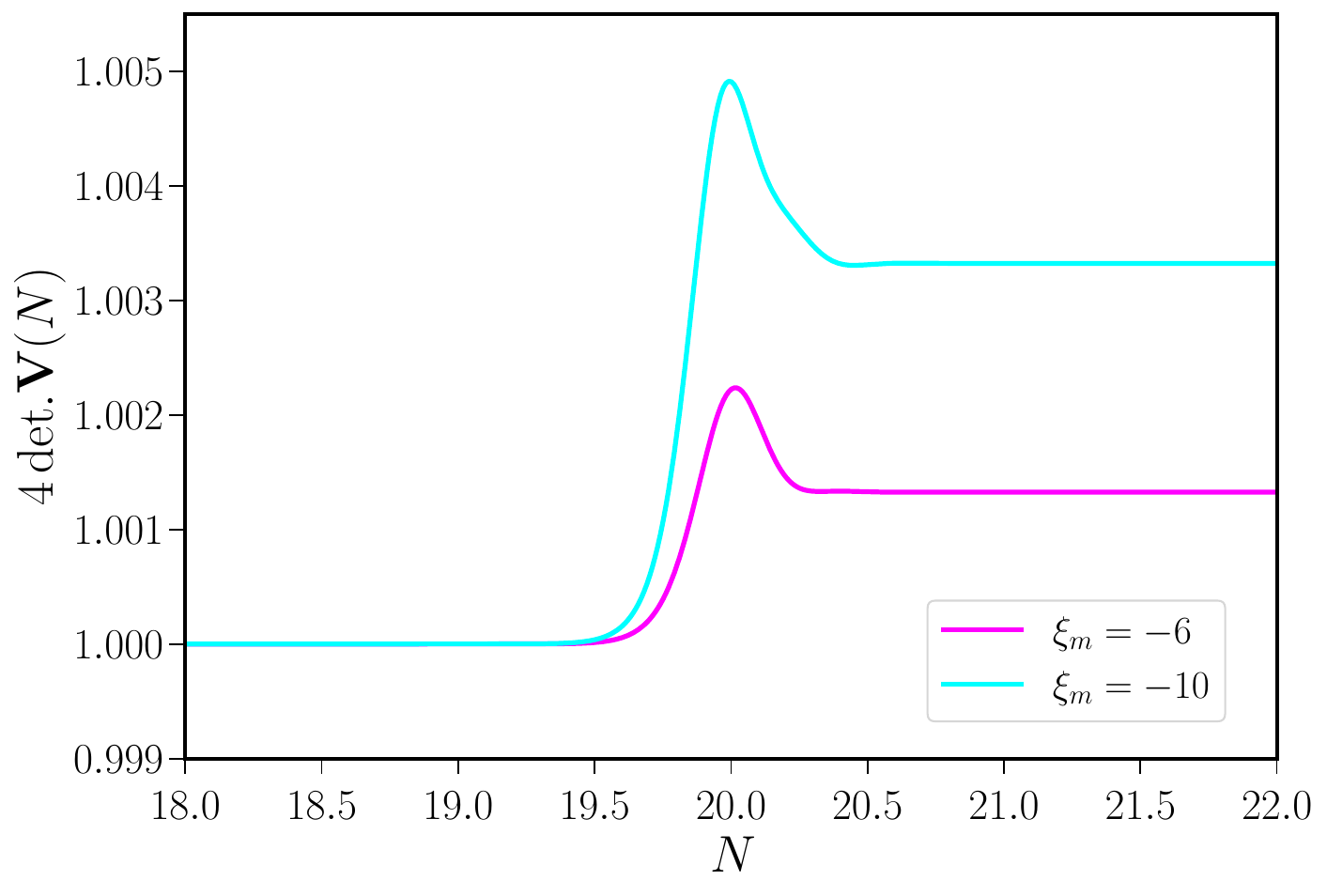}~~
\includegraphics[width=0.45\linewidth]{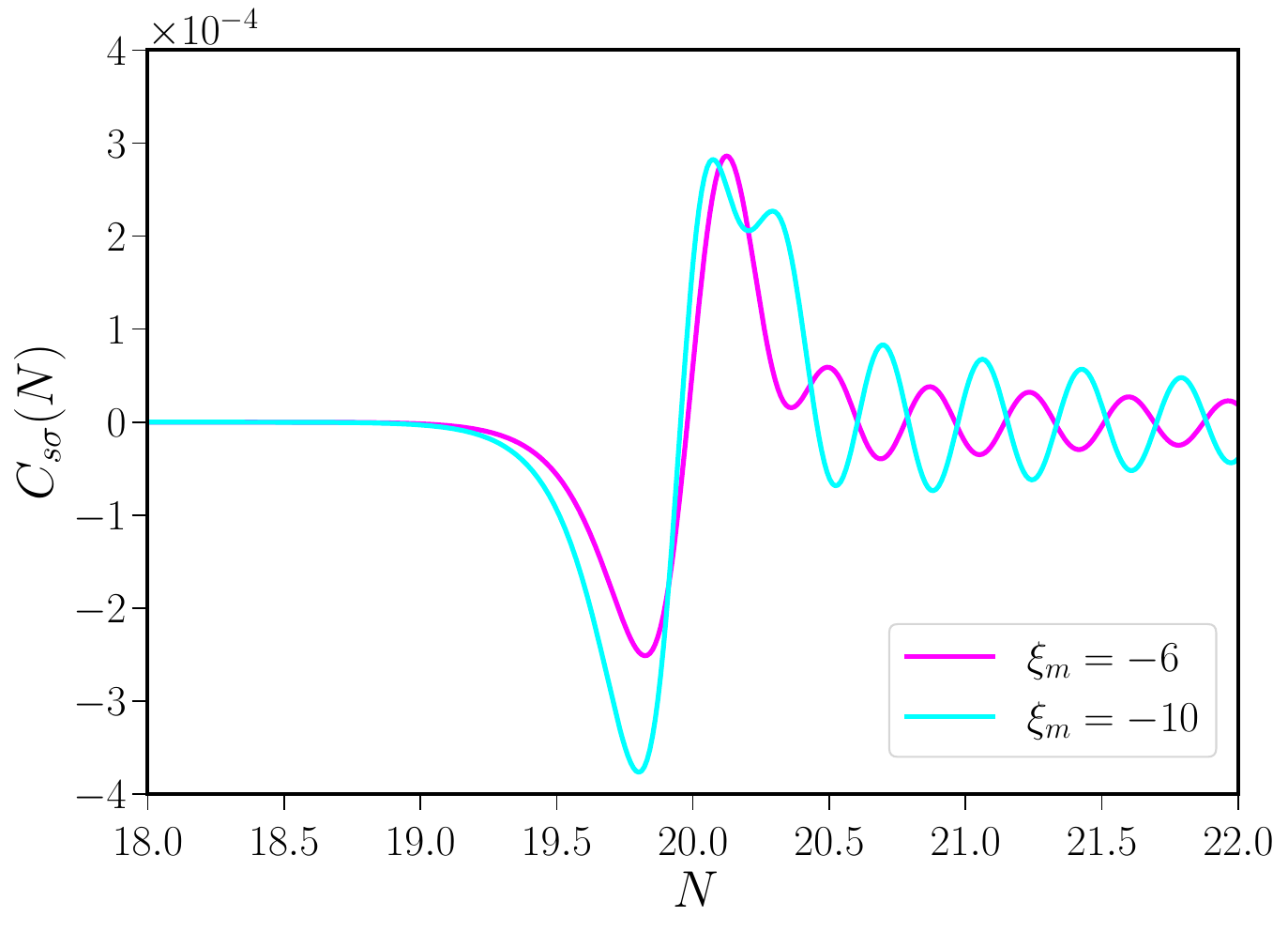}
\caption{Evolution of the quantities ${\rm det}.\bm{V}$ and $C_{s \sigma}$ with the e-folding parameter $N$ has been depicted. For these plots we have chose the coupling function $\xi$ to peak at $N_0=20$, and have taken the wave number to be $k=10^{-4}{\rm Mpc^{-1}}$.}
\label{fig:VCss}
\end{figure}

The coupling function for the present case is described by the same function as in \ref{eq:xibaH}, but with a different value for $N_{0}$, such that $N_{0}<N_{\rm p}$, with $N_{\rm p}$ being the e-folding parameter where the pivot scale exit the Hubble horizon. The structure for the coupling function with the e-folding parameter $N$, for such a choice for $N_{0}$, has been presented in \ref{fig:xibaH-N0-35} with different values for the amplitude $\xi_{\rm m}$. For this variation of the coupling function, the evolution of the determinant of the covariance matrix and the quantity $C_{s\sigma}$ has been presented in \ref{fig:VCss}. As evident, turning on the coupling leads to a non-trivial evolution of $\textrm{det}.\bm{V}$, in particular, it is enhanced and then attains a constant value after the coupling function is turned off. The quantity $C_{s\sigma}$, on the other hand, depicts an oscillatory behaviour even after the coupling function has vanished. This suggests that any sharp deviation from the background trajectory can result in oscillatory features in the power spectrum. Note that the choices of the parameters in this scenario is following the work \cite{Achucarro:2010da}. As in the case of the nearly scale-invariant power spectrum discussed earlier, Although there is an increase in the evolution of $\textrm{det}.\bm{V}$, it decreases due to the change in the sign in $C_{s \sigma}$. Additionally, the oscillations in  $C_{s \sigma}$ are not visible in $\textrm{det}.\bm{V}$ since the coupling parameter decays rapidly during that time period.

\begin{figure}[]
\centering
\includegraphics[width=0.45\linewidth]{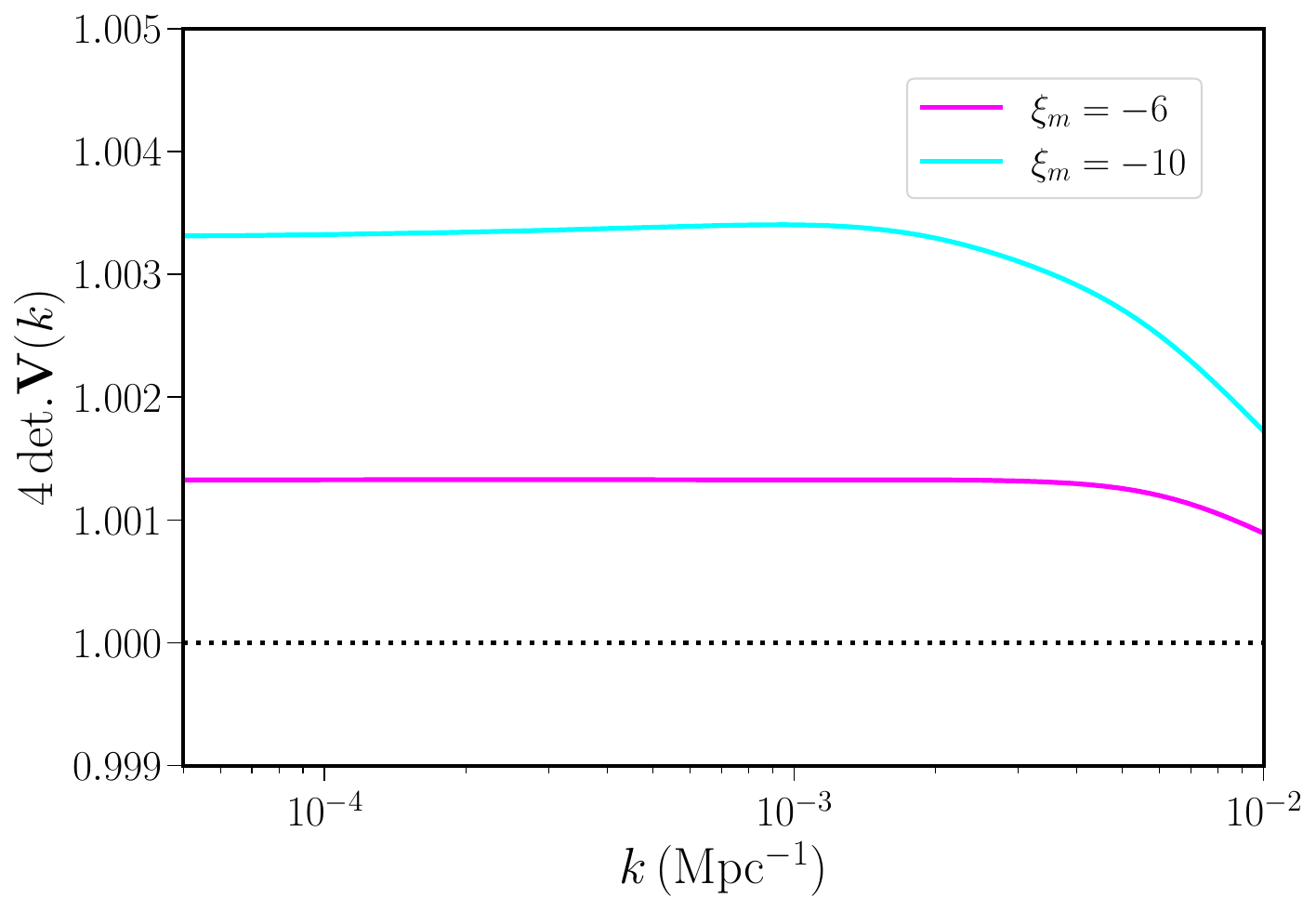}~~
\includegraphics[width=0.45\linewidth]{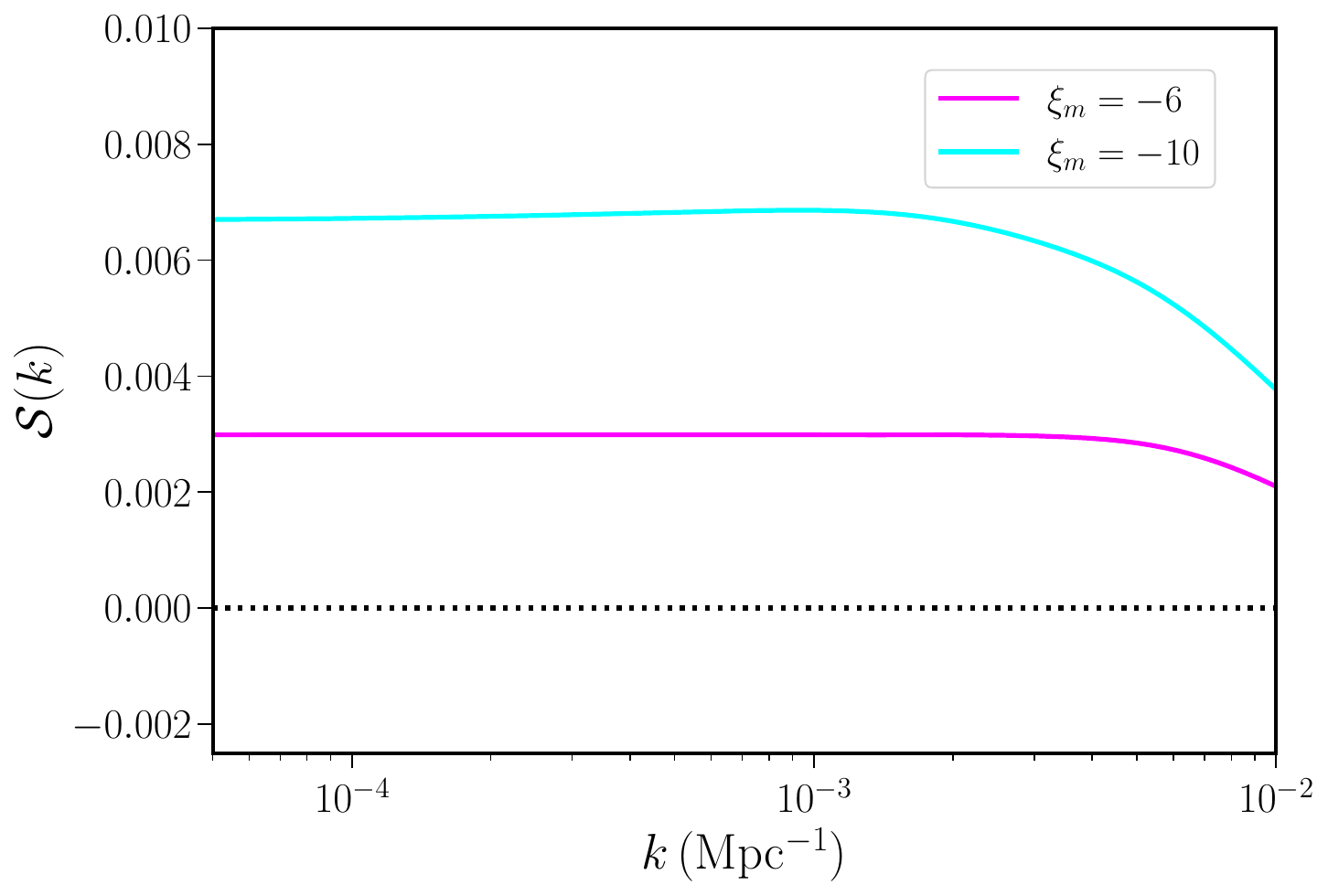}
\caption{Spectra of ${\rm det}.\bm{V}$ and entanglement entropy $\mathcal{S}$ have been presented, after the turning, for different values of the amplitude $\xi_m$ for the coupling function. In this model, the amplitude and the form of $\xi$ lead to oscillatory features in the power spectrum of the curvature perturbations. It is important to note that both ${\rm det}.\bm{V}$ and $\mathcal{S}$ depict different behaviour for the single-field and two-field models. These features can be used to distinguish between various early universe cosmological models of our universe.}
\label{fig:delta-2f-1f-EFT}
\end{figure}

We have also plotted the spectra associated with the $\textrm{det}.\bm{V}$ and the entanglement entropy $\mathcal{S}$ over the CMB scales in \ref{fig:delta-2f-1f-EFT}. It is important to note that $\textrm{det}.\bm{V}$ is equal to one and therefore entanglement entropy $\mathcal{S}$ is zero in single-field models even though it can produce similar features in the curvature power spectrum. While for two-field models of inflation both $\textrm{det}.\bm{V}$ and $\mathcal{S}$ depict non-trivial evolution with the e-folding parameter. This explicitly demonstrates that two distinct models of inflation, producing identical power spectra, can have significantly different quantum measures. Therefore, these quantum measures, e.g., purity (which is simply the inverse of the determinant of the covariance matrix) and entanglement entropy, can be used to distinguish single-field model of inflation from the multi-field models. Thus, detectability of these quantum measures from the Planck data or, using data from future missions will immensely help in identifying the cosmological model describing the early universe, be it single or multi-field models. However, it is worth noting that the direct measurement of the determinant of the covariance matrix, $\textrm{det}.\bm{V}$, is not achievable solely through current observations. The determination of $\textrm{det}.\bm{V}$ necessitates the consideration of momentum correlations and curvature perturbation-momentum correlations, whose measurement will require future generations of CMB missions. For completeness, we will also present the corresponding results for the multi-field model in the bouncing scenario. 
\begin{figure}[]
\centering
\includegraphics[width=0.45\linewidth]{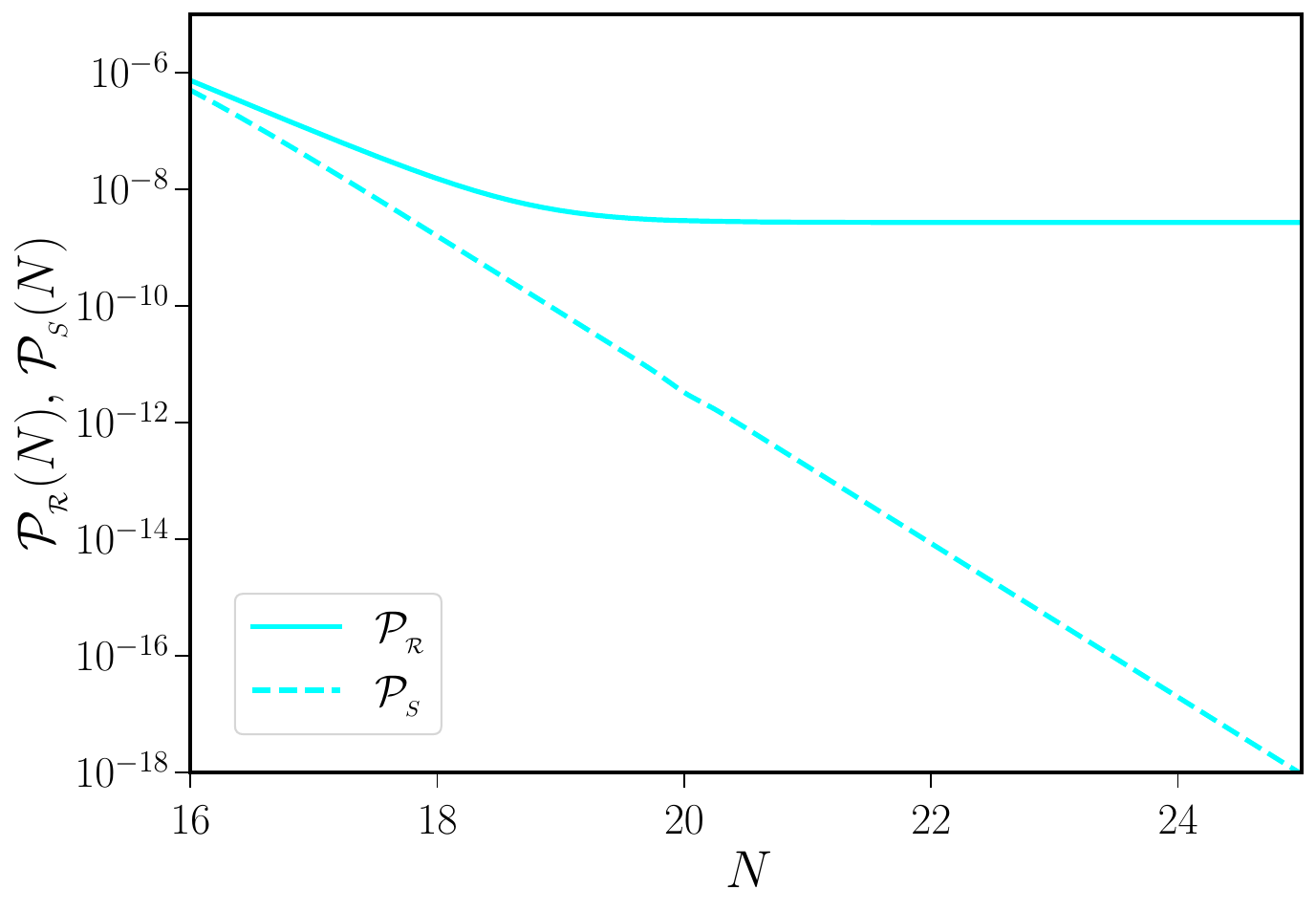}~~
\includegraphics[width=0.45\linewidth]{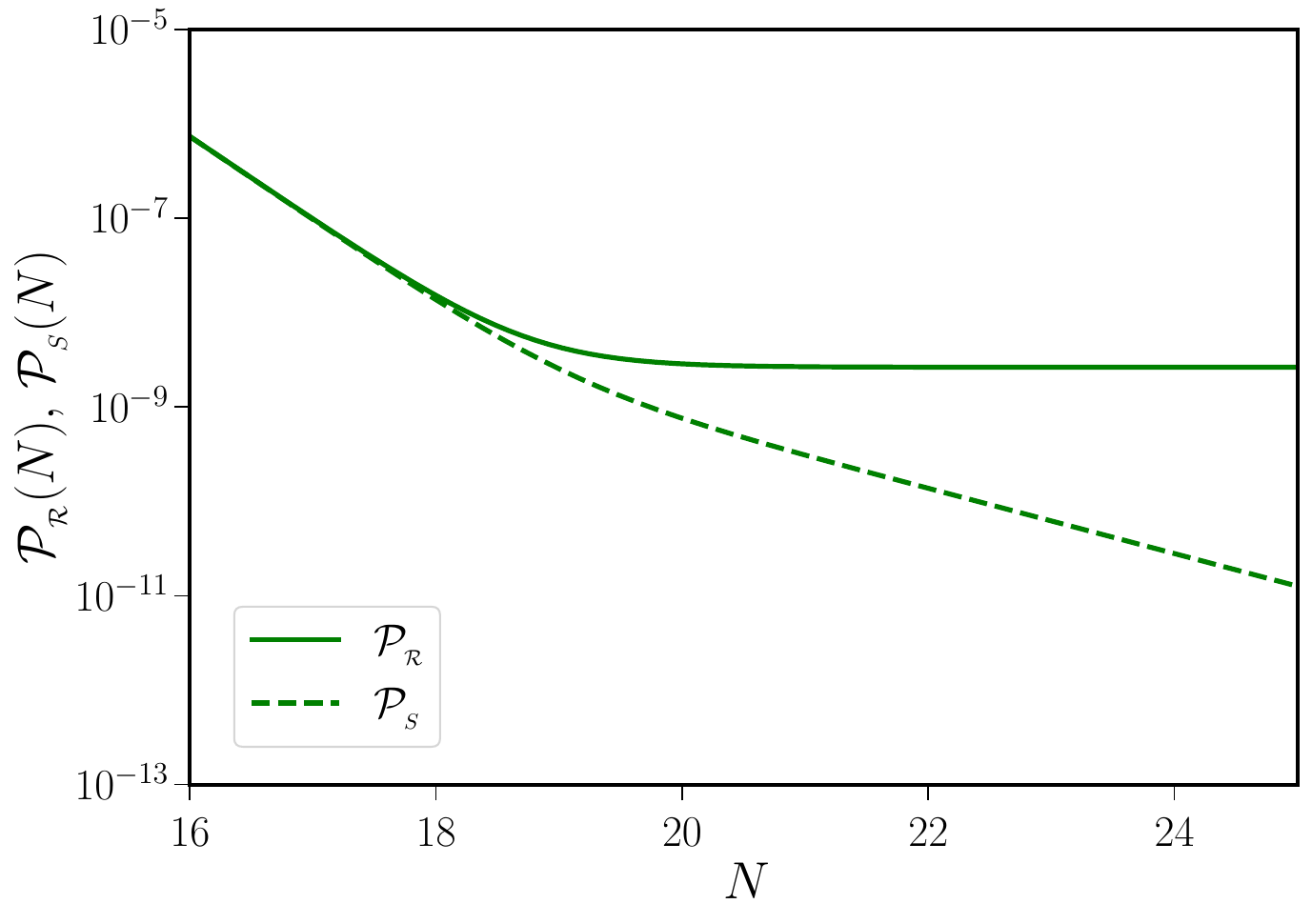}
\caption{We illustrate the evolution of the power associated with the curvature and the isocurvature perturbations as a function of the e-folding parameter $N$. The plot on the left showcases the evolution of the power in the curvature and the isocurvature perturbations when the coupling function $\xi$ reaches its peak at $N_0=35$, while the figure on the right side exhibits the evolution of the respective power spectrum when $\xi$ peaks at $N_0=20$. These particular selections correspond to the models elaborated in \ref{sec:model-SI} and \ref{sec:model-features}, respectively. In both the cases, we have set $\xi_m=-10$ and have chosen the wave number to be $k=10^{-4}{\rm Mpc^{-1}}$. It is evident from the figures that the isocurvature perturbations decay over time and become significantly smaller compared to the curvature perturbations towards the end of the inflation.}
\label{fig:PR-N}
\end{figure}

We also present the evolution of power related to curvature and isocurvature perturbations in \ref{fig:PR-N}. As observed in the figures, isocurvature perturbations exhibit a noticeable decay over time, becoming substantially smaller compared to the curvature perturbations towards the conclusion of the inflationary period. These results provide evidence that the considered models produce only the adiabatic perturbations at the end of inflation, aligning with the observations by PLANCK~\cite{Ade:2015lrj}.

\section{Two-field ekpyrosis}\label{ekpyrosis}

We have already shown how the quantum features in the perturbations of the scalar fields in the early universe can be used to distinguish between single and multi-field models of inflation. Though, we have shown this explicitly for the two-field model, the formalism can be easily generalized and the effects will possibly be more enhanced for multi-field models. However, all of these conclusions were arrived in the context of inflationary cosmology. For completeness, in this section, we apply our formalism for two-field bouncing models as well, namely for two-field ekpyrotic cosmology. 

It turns out that nearly scale-invariant power spectrum for curvature perturbation can also be generated in the ekpyrotic bounce scenario. This is achieved through the conversion of the isocurvature/entropic perturbation to curvature perturbation (for related discussions see \cite{Li:2013hga, Ijjas:2014fja, Fertig:2016czu, Raveendran:2018yyh}), and as discussed in the previous section, the conversion of isocurvature perturbations to curvature perturbations can occur  when there is a change in the trajectory of the field in the field space, e.g., due to the coupling function $\xi$. In the context of ekpyrotic bouncing cosmology, the conversion of isocurvature/entropic perturbation to curvature perturbation takes place in the contracting phase, i.e., before the bounce. Following which, we investigate the evolution of quantum signatures during the ekpyrotic contracting phase. As we know the change in the background trajectory can be represented by the parameter $\xi$, as seen in \ref{eq:xibaH}, in the case of two-field inflation. Here, we additionally need to specify the entropic mass $\mu_s^2$, which is responsible for the shape of the curvature power spectrum. Following \cite{Raveendran:2018yyh}, the entropic mass is chosen to be,
\begin{equation}
\mu_s^2= C \, \frac{\left(6-\lambda^2 -\lambda\,\mu\right)\lambda\,\mu}{\left(\lambda^2-2\right)^2}{\rm exp}\left[-\lambda^2 \, N\right]-\left(\frac{\xi}{2 \,a}\right)^2~,
\end{equation}
where $C$, $\lambda$ and $\mu$ are constants and need to be chosen based on the structure of the curvature power spectrum, in particular, its amplitude and slope. With this choice for the entropic mass function, along with the choice of the coupling function $\xi$, we have determined the evolution of the determinant of the covariance matrix with the e-folding parameter. 

\begin{figure}[h!]
\centering
\includegraphics[width=0.45\linewidth]{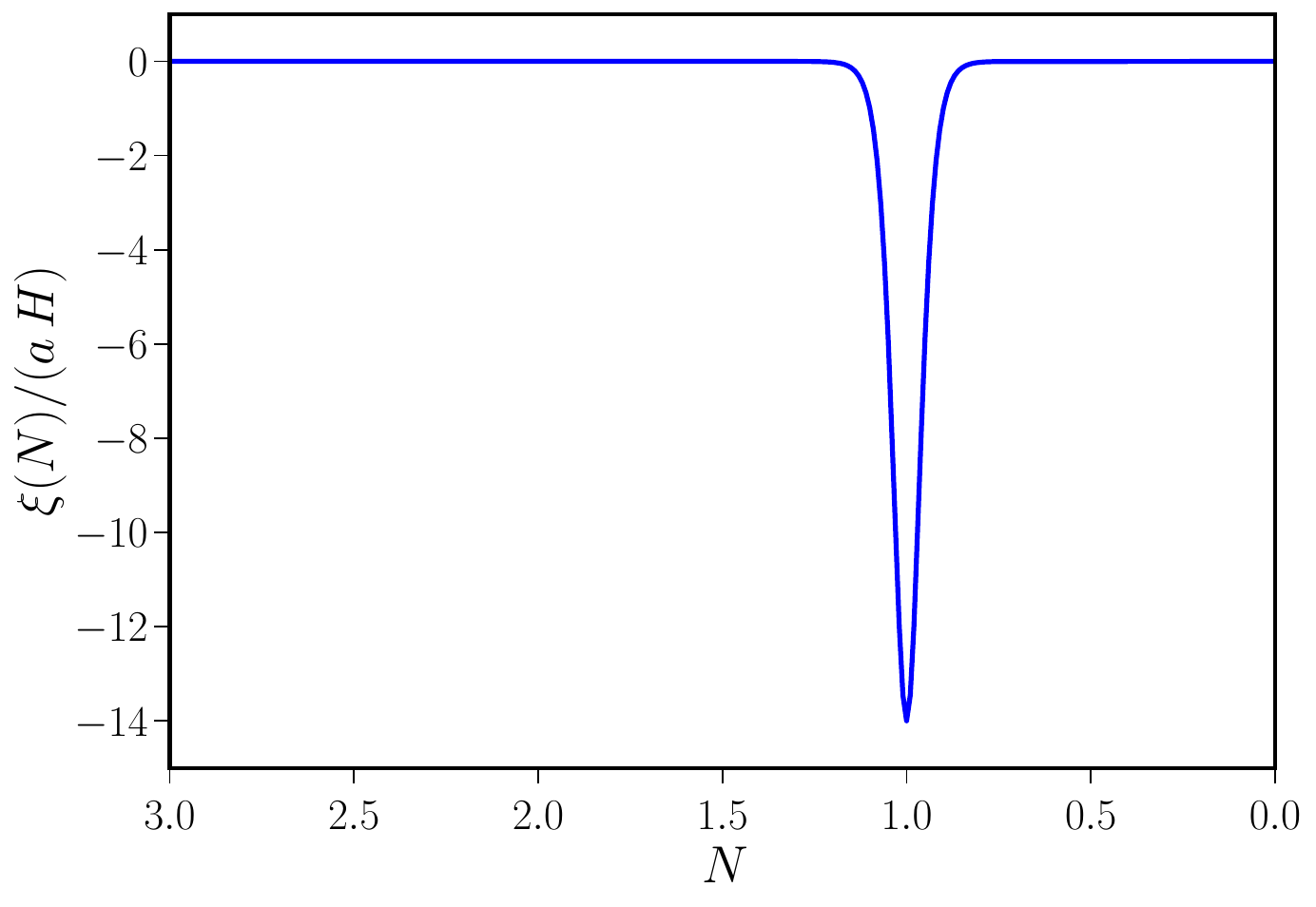}~~
\includegraphics[width=0.45\linewidth]{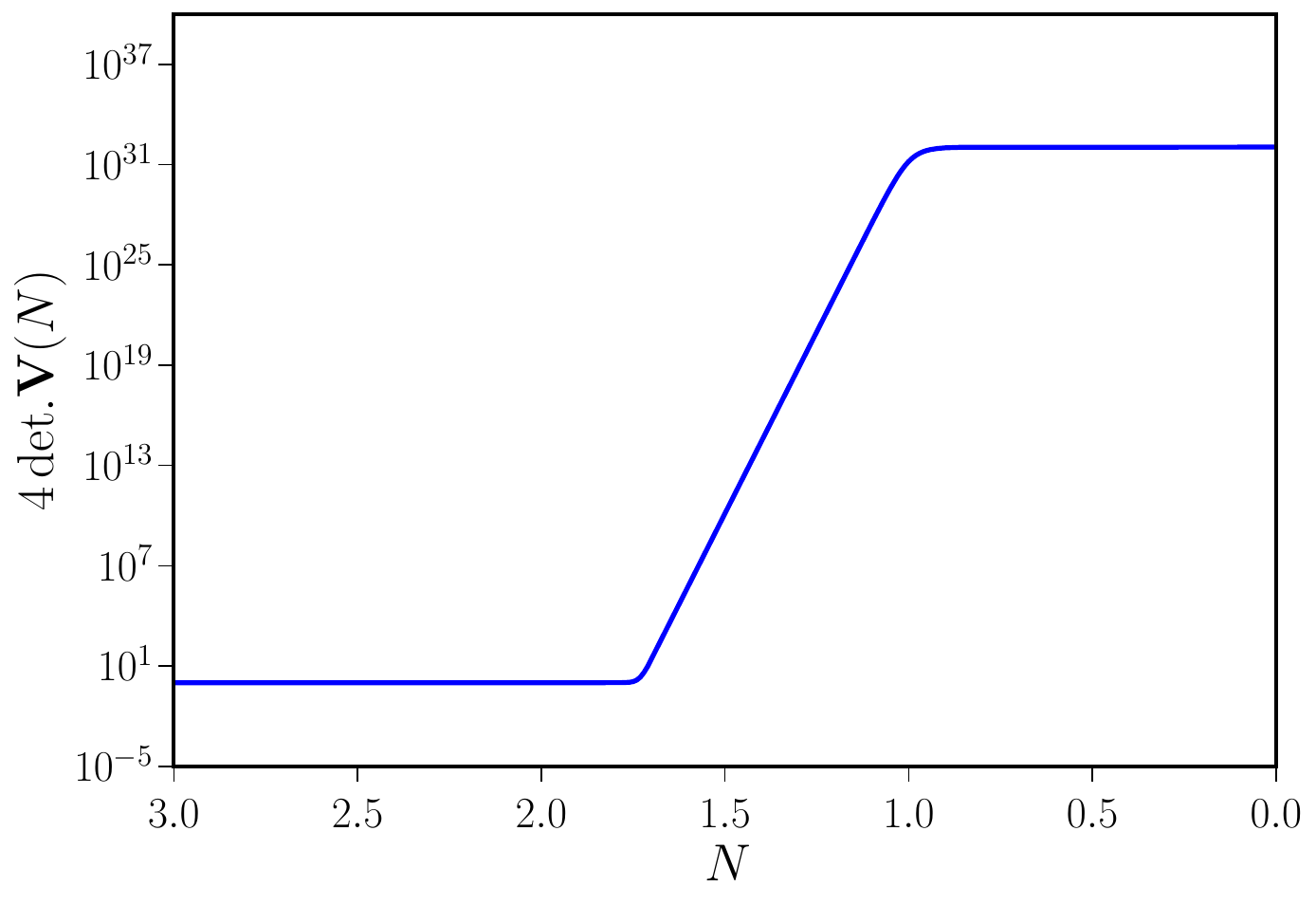}
\caption{Evolutions $\xi$ and ${\rm det}.\bm{V}$ during ekpyrotic contracting phase. It is known that the amplitude of the curvature power spectrum is used to determine the value of $C$, which is equal to $769.42$. Additionally, the values of $\lambda^2$ and $\mu$ are chosen based on the shape of the power spectrum, with $\lambda^2=20$ and $\mu=4.553$.}
\label{fig:etap}
\end{figure}

The results of the evolution of the $\textrm{det.}\bm{V}$, along with the evolution of the coupling function, have been plotted during the ekpyrotic contracting phase in \ref{fig:etap}. As we observed in the context of inflation, here also $\textrm{det}.\bm{V}$ increases due to the interaction term. As a consequence, as in the case of multi-field inflation, this results into a decrease in the purity and increase in the entanglement entropy of the reduced system involving the curvature perturbation alone. Moreover, the evolution with time, as well as the value achieved by $\textrm{det}.\bm{V}$ at during the contracting phase is significantly different than what is expected from inflation. Since all the other quantum measures are related to the determinant of the covariance matrix, we expect that the quantum measures in the ekpyrotic phase will be significantly different from that of the inflationary counterpart. So one can conclude that even if different models, both with the context of inflation and bounce, share the same observable power spectrum of curvature perturbations, their quantum signatures, such as purity, entanglement entropy, and quantum discord can have very different values. 
\section{Discussion and concluding remarks}

In this work, we have examined the evolution of the quantum signatures of the curvature perturbations in the multi-field models of inflation. We have used mathematical measures such as the behavior of the Wigner function, purity and entanglement entropy (arrived at by integrating the degrees of freedom associated with the isocurvature/entropic perturbations) or, equivalently, quantum discord. Even though our results are general and apply to early universe consisting of an arbitrary number of scalar fields, for illustration, we have specifically studied two-field models in both inflation and bouncing cosmology. In particular, we have discussed two scenarios --- (a) the curvature perturbations do not interact with the entropic perturbations, before the relevant modes exit the Hubble horizon, and (b) the curvature and entropic perturbations interact, even before the curvature perturbation could exit the Hubble horizon. In the case where entropic perturbation does not leave any imprint on the observable power spectrum of curvature perturbations, it affects the quantum measures holding information about various properties of the quantum state. Thus quantum properties, if observed, can predict the existence of entropic perturbation, through the coupling with the curvature perturbation.

A similar story is repeated when we study models generating a power spectrum with oscillatory features, due to interaction between curvature and entropic degrees of freedom, even before the curvature modes could exit the Hubble horizon. Though we can mimic the power spectrum of these two-field models by an effective single-field description, the quantum measures are largely different. For example, the determinant of the covariance matrix is exactly equal to one for single-field model, but deviates significantly from zero in the case of multi-field models. Thus even in this case, a handle on the quantum measures can actually distinguish single-field and multi-field models of inflation. 

The above result does not stop at inflationary paradigm alone, but transcends to bouncing models with multiple fields, e.g., in the context of ekpyrotic bounce. Even there, the specific values of quantum measures can distinguish it from single field models, though they will have indistinguishable curvature power spectrum. Thus we can safely argue that quantum measures are the appropriate tool to distinguish multi-field models with interactions from single-field models, be it inflation or, bounce. In other words, different inflationary/bouncing models for the early universe cosmology may lead to similar results, as long as large scale structure of the universe is considered, but may still differ in other ways, such as in their quantum signatures. If such quantum measures are detected in the near future, they can be used to compare the predictions of these models and potentially distinguish between them. Although we have used the specific case of two-field models, we believe the general formalism developed here can be applied to other multi-field models as well. One may question whether certain variables could yield identical values for quantum measures, particularly ${\rm det}.\bm{V}$, in both inflationary and bouncing scenarios. If such a scenario exists, distinguishing between these models based solely on quantum measures may prove challenging. We anticipate that the entanglement entropy calculated using the same set of variables and states will differ between inflationary and bouncing models. This difference arises from the evolution of ${\rm det}.V$, which is influenced by the interaction between curvature and isocurvature perturbations (see \ref{eq:detV-de-2f}). Given that this interaction varies between the two paradigms, the entanglement entropy will likewise differ. Consequently, it should be feasible to differentiate between inflationary and bouncing models of cosmology. We intend to explore more on this direction in our future work.

There are also several future directions to embark upon, e.g., we have established that recoherence (or, the restoration of purity) is possible with the interaction between the curvature and isocurvature/entropic perturbations, for certain parameter selections. It would be interesting to pursue this result further and observe if the perturbations underwent repeated phases of classicalization and then again become quantum. Moreover, it would interesting if some avenue to observe the imprint of these quantum measures on the CMB physics can be studied. The detection of such imprints can eventually distinguish between various models of inflation and bounce, as they have very different estimation and evolution of quantum measures. We acknowledge that, in this work, we investigate the quantum measures of cosmological perturbations in Fourier space. However, it is important to note that observations are typically conducted in real space. Therefore, one should also compute the quantum measures of the quantum field in real space, as discussed in \cite{Martin:2021qkg}. Moreover, in the context of single-field models, the findings of \cite{Martin:2021qkg} reveal that although there may be a non-negligible presence of entanglement entropy or quantum discord in real space, these quantities are significantly suppressed. However, it is worth noting that these measures can still be significant in Fourier space. Then, it is interesting to explore how these findings differ in multi-field models. We anticipate that although quantum measures are significantly suppressed, differences between single-field and multiple-field scenarios will persist. Additionally, it is interesting to examine how the quantumness obtained in Fourier space can be translated into measurements of CMB, which mostly rely on observables in real space. In this context, one can expect that the detection of the difference in quantumness between single and
multi-field models can be challenging in CMB, but the difference will be present. Further investigation into this aspect is also planned as part of future projects.


\section*{Data availability}

Data sharing is not applicable to this article as no data sets were generated 
or analyzed during the current study.

\section*{Acknowledgements}

The authors wish to thank Krishnamohan Parattu and L. Sriramkumar for interesting discussions. RNR is supported by post-doctoral fellowship from the Indian Association for the Cultivation of Science, Kolkata, India. Research of SC is funded by the INSPIRE Faculty fellowship from DST, Government of India (Reg. No. DST/INSPIRE/04/2018/000893).

\appendix

\labelformat{section}{Appendix #1}
\labelformat{subsection}{Appendix #1}
\labelformat{subsubsection}{Appendix #1}
\section*{Appendices}

\section{Evolution in the Heisenberg picture}

In the main text, we have discussed the evolution of the primordial perturbations in the Schr\"{o}dinger picture. Here we show how the evolution can be obtained in the Heisenberg picture as well, by promoting the fields to operators. In terms of creation and annihilation operators, the operators associated with the Mukhanov-Sasaki variables and the conjugate momenta can be written as,
\begin{subequations}\label{eq:Heisenberg-expn}
\begin{align}
\hat{v}_m =\sum_{n}\left( f_{m n } \hat{a}_n + f_{m n}^* \hat{a}_n^\dagger\right)\equiv \bm{F}^{\rm T} \, \hat{\bm{A}} + \bm{F}^{\rm T*} \hat{\bm{A}}^\dagger~,
\\
\hat{p}_m=\sum_{n}\left( \pi_{nm} \hat{a}_n + \pi_{nm}^* \hat{a}_n^\dagger\right)\equiv \bm{\Pi}^{\rm T} \, \hat{\bm{A}} + \bm{\Pi}^{\rm T*} \hat{\bm{A}}^\dagger~,
\end{align}
\end{subequations}
where, the elements of the matrices $\bm{F}$ and $\bm{\Pi}$ are the mode functions associated with the respective operators. From the Heisenberg equations of motion, one can see that the mode functions, $\bm{F}$ and $\bm{\Pi}$ obey the classical equations as
\begin{subequations}\label{eq:F-Pi-eqns}
\begin{align}
\bm{F}' &= \bm{\Pi} + \bm{A}\,\bm{F}~,
\\
\bm{\Pi}' &= -\bm{\Pi}\,\bm{A} -2\,\bm{F}\,\bm{B}~.
\end{align}
\end{subequations}
From the commutation relations between the creation and annihilation operators, 
\begin{equation}\label{eq:cc}
\left[\hat{a}_m,\hat{a}_n\right]=\delta_{mn}~,
\end{equation}
we can obtain the Wronskian associated with the mode functions as 
\begin{equation}\label{eq:wronskians}
\bm{F}\bm{\Pi}^\dag - \bm{\Pi} \bm{F}^\dag=i\, \bm{I}~.
\end{equation}
Therefore, the two-point correlation functions associated with the above fields can be expressed in terms of the mode functions as,
\begin{subequations}\label{eq:F-Pi-correlations}
\begin{align}
\left \langle  \hat{v}_{m}\hat{v}_{n}\right \rangle =& \bm{F}^\dag \bm{F}~,
   \\
\left \langle  \hat{p}_{m}\hat{p}_{n}\right \rangle =& \bm{\Pi}^\dag \bm{\Pi}~,
\\
\left \langle  \hat{v}_{m}\hat{p}_{n}+\hat{p}_{n}\hat{v}_{m}\right \rangle= & \bm{F}^\dag \bm{\Pi}+\left(\bm{F}^\dag \bm{\Pi}\right)^\ast~.
\end{align}
\end{subequations}
In the Heisenberg formalism, one can solve the equations of motion provided in \ref{eq:F-Pi-eqns} and obtain the correlation functions using \ref{eq:F-Pi-correlations}. Recall that, in the Schr\"{o}dinger formalism, one also solves for the matrix $\bm{\Omega}$, appearing in the Schr\"{o}dinger wave function, using \ref{eq:de-Omega} and obtain the correlation function from \ref{eq:correlations}. The situation in the Heisenberg picture can be obtained by identifying the relation between $\bm{\Omega}$ and the mode functions defined in \ref{eq:Heisenberg-expn}. In other words, this is done by using \ref{eq:de-Omega-matrix} to identify $\Omega$ as
\begin{equation}\label{eq:Omega-modefunctions}
 i\,\bm{\Omega}=\left(\bm{F}^{-1}\, \bm{\Pi}\right)^*~, \qquad {\rm or}\qquad i\, \bm{\Omega}=\bm{F}^{-1} \, \bm{\Pi}~.
\end{equation}
This provides the connection between the evolution in the Schr\"{o}dinger and the Heisenberg picture. 



\bibliography{mybibliography-cosmology.bib,mybibliography-quantum.bib}

\end{document}